\documentstyle[preprint,eqsecnum,prd,aps]{revtex}
\begin{document}
% states and fields

%%%%Start of Text%%%%%%%%%%%%%%%%%%%%%%%%%%%%%%%%%%%%%%%%%%%%%%%%%%%%%%%%%%%%
%\rightline{
\preprint{
\vbox{
\halign{&##\hfil\cr
	& ANL-HEP-PR-95-36 \cr
	& December 15, 1995 \cr}}
}
%}
\title{
Analytic Calculation of Prompt Photon plus Associated Heavy Flavor at
Next-to-Leading Order in QCD}
\author{Edmond L. Berger and L. E. Gordon}
\address{High Energy Physics Division, Argonne National Laboratory,
	Argonne, IL 60439}
\maketitle
\begin{abstract}
Contributions through second order, $O(\alpha ^2_s)$, in perturbative quantum
chromodynamics are calculated analytically for inclusive associated production
of a prompt photon and a charm quark at large values of transverse momentum in
high energy hadron-hadron collisions.  Seven partonic subprocesses contribute
at order $\alpha^2_s$.  We find important corrections to the lowest order,
$O(\alpha_s)$, subprocess $c g \rightarrow \gamma c$.  We demonstrate to what
extent data from $p +\bar{p}\rightarrow \gamma + c + X$ may serve to measure
the charm quark density in the nucleon.

%\end{abstract}
\vspace{0.2in}
\end{abstract}
\narrowtext
\section{Introduction}

Because photons couple in point-like fashion to quarks, observation,
among the final-state particles in a high energy collision, of photons carrying
large values of transverse momentum provides an incisive probe of the short
distance hadron dynamics of the collision.  This fact explains the substantial
theoretical and experimental interest shown in studies of the cross section
for production of photons at large angles in hadron-hadron and
lepton-hadron scattering and in electron-positron annihilation processes.  At
stake are precise tests of the theory of perturbative quantum chromodynamics
(QCD) and use of data to determine properties of the relativistic proton such
as the momentum distribution of its constituent gluons and quarks.  Discovery
of the charm quark and, later, of the bottom quark stimulated interest in the
dynamics of their relatively copious production in high energy interactions of
hadrons.  Recent experimental advances now offer the possibility of studies of
the associated production of a photon $(\gamma)$  carrying large transverse
momentum along  with a heavy quark $(Q)$ whose transverse momentum balances a
substantial portion of that of the photon.\cite{cdf}  In this paper, we
report a fully analytic next-to-leading order QCD calculation of the
two-particle inclusive distribution for prompt photon plus associated heavy
flavor production at large values of transverse momentum, with specification
of the momentum variables of both the final prompt photon and the final heavy
quark.  These results should facilitate further experimental tests of
correlations inherent in the QCD matrix elements and provide a means for
measuring the charm quark density in the nucleon.

Although a qualitative description may be obtained from lowest-order
perturbation theory, more precise predictions of the momentum distribution
for the inclusive production a heavy quark (or antiquark) require perturbative
calculations that extend to higher order.\cite{qnlo}  Likewise, perturbative
QCD calculations of inclusive and isolated prompt single photon production
are available.\cite{gamnlo1,gorvogel,berqiu}  At the level of two-particle
inclusive final states, next-to-leading order QCD calculations have been done
for ${\gamma\gamma}$ production\cite{aur2,boboo}, for $\gamma$-hadron
production \cite{aur3} and for $\bar{Q}Q$ correlations.\cite{correl}  The
cross section for the production of two hadronic jets has been studied at
$O(\alpha ^3_s)$ by several authors.\cite{sdellis}
Constraints on the charm and strange quark densities from data on intermediate
vector-boson production are discussed in Ref.~\cite{ELBW}.

For values of transverse momentum $p^Q_{T}$ of the heavy quark
significantly larger than the mass $m_Q$ of the heavy quark, the cross section
for the two-particle inclusive reaction  $p +\bar{p}\rightarrow \gamma + Q + X$
may be calculated from the leading order QCD subprocess, the quark-gluon
Compton process,  $g + Q \rightarrow \gamma + Q$.  This subprocess is of first
order in the strong  coupling strength $\alpha_s$.  The cross section is
obtained as a convolution of the hard-scattering QCD matrix with probability
distributions that specify the initial gluon and heavy quark constituent
momentum densities in the incident hadrons, $p$ and $\bar{p}$.  At
next-to-leading order in QCD, several subprocesses contribute to the $\gamma +
Q$ final state:
\begin{mathletters}\label{eq:1}
\begin{eqnarray}
g &+ Q \rightarrow g + Q + \gamma\label{eq:11}\\
g &+ g \rightarrow Q +\bar{Q} + \gamma\label{eq:12}\\
q &+ \bar{q} \rightarrow Q +\bar{Q} + \gamma\label{eq:13}\\
q &+ Q \rightarrow  q + Q + \gamma\label{eq:14}\\
\bar{q} &+ Q \rightarrow \bar{q} + Q + \gamma\label{eq:15}\\
Q &+ \bar{Q} \rightarrow Q + \bar{Q} + \gamma\label{eq:16}\\
Q &+ Q \rightarrow Q + Q + \gamma\label{eq:17}
\end{eqnarray}
\end{mathletters}
For computation of the cross section for $\bar{Q}$ production, the set
of next-to-leading order subprocesses is obtained from those of
Eq.~(\ref{eq:1}) after
replacement of the initial $Q's$ by $\bar{Q}'s$
in Eqs.~(\ref{eq:11}), (\ref{eq:14}), (\ref{eq:15}), (\ref{eq:17}).
We note that for values of $p^Q_{T}$ that are comparable to or
less than $m_Q$ there would be no $O(\alpha_s)$ subprocess, and the proper hard
scattering expansion would entail only the subprocesses of Eqs.~(\ref{eq:12})
and (\ref{eq:13}).  For the remainder of this paper, we limit
ourselves to charm production, and we work with the massless Q approximation,
$m_c = 0$.

We are interested ultimately in the fully differential two-particle
inclusive cross section,  $E_\gamma E_Qd\sigma/d^3p_\gamma d^3p_Q$, where
$(E,p)$ represents the four-vector momentum of the $\gamma$ or $Q$.  For each
subprocess listed in Eq.~(\ref{eq:1}), this calculation requires integration
of the
momentum of the unobserved final parton ($g$, $\bar{Q},q$, or $\bar{q}$) and
over the initial parton momentum densities. Collinear singularities are handled
analytically by dimensional regularization and absorbed into initial-state
parton momentum densities or final-state fragmentation functions.  To make
the analytic calculation tractable, we
chose to work in terms of the transverse momentum of the final $\gamma$,
$p^{\gamma}_{T}$, and the ratio of the heavy quark and photon transverse
momenta:
\begin{equation}
z = - {{p^Q_{T}. p^{\gamma}_{T}}\over {(p^\gamma_{T})^2}}.  \label{eq:zdef}
\end{equation}

To warrant use of perturbation theory (and the massless $Q$ approximation), we
limit our considerations to $z > 0$ and $p^\gamma_{T} > 10$ GeV.  The results
should be applicable quantitatively for $p^c_{T} \gg m_c$. The
distribution in $z$ from the leading order subprocess
$g + Q \rightarrow \gamma + Q$ is peaked sharply at $z = 1$
(a $\delta(1-z)$ function in the naive collinear initial parton
approximation).  The next-to-leading order processes alter the
size of this sharp peak and produce a broad distribution above and below
$z = 1$.

Contributions to hard photon production from long-distance quark to photon
and gluon to photon fragmentation processes have been
emphasized theoretically,\cite{field} parametrized phenomenologically in
leading order,\cite{frag1} and evolved in next-to-leading
order.\cite{frag2,frag3} These terms
may account for more than half of the
calculated inclusive single photon cross section at modest values of transverse
momentum at the Fermilab Tevatron collider.  Because of our kinematic
restriction $z > 0$, there will be no contributions to the final cross
section from $Q \rightarrow \gamma$ fragmentation, where $Q$ is the observed
quark/anti-quark, from among the subprocesses
in Eq.~(\ref{eq:1}).  On the other hand, fragmentation of the unobserved final
parton into a photon in subprocesses (1.a-g) will contribute to the cross
section and produce photons that carry $p_T$ less than that of $p^Q_{T}$,
mostly populating the region $z > 1$.  Photons originating through
fragmentation are likely to emerge in the neighborhood of associated hadrons.
An experimental
isolation restriction is needed before a clean identification can be made of
the photon and a measurement made of its momentum.  Isolation reduces the size
of the observed fragmentation contribution.  To represent the effects of
isolation, we should use fragmentation functions defined with a cone size.
Photon
isolation complicates the theoretical interpretation of results, however, since
it threatens to upset the cancellation of infra-red divergences in perturbation
theory.\cite{berqiu} In this paper, we calculate the contributions from photon
fragmentation at leading order only, and, except for one illustrative figure,
we neglect the isolation requirements.

After integration over the longitudinal momentum of the heavy quark, we
present our results in terms of the cross section
$d\sigma/dp^\gamma_{T} dy^\gamma dz$. Here, $y^\gamma$ represents the
rapidity of the $\gamma$.  Our
desire to perform a fully analytic calculation restricts our ability to provide
a more differential cross section in this paper (i.e., a cross section also
differential in $y^Q$).  In a later more detailed paper, we will present such
results obtained from a versatile combination of analytic and Monte Carlo
techniques.\cite{moncarlo}  In that method, selections may be made on several
variables and photon isolation restrictions are easier to impose.  An earlier
theoretical paper addresses prompt photon plus associated charm
production at large values of transverse momentum, as we do here, but
our analysis differs from that of Ref.~\cite{strvog}.
The calculation of the photon-plus-charm cross section in
Ref.~\cite{strvog} is done in lowest order while ours is done at
next-to-leading order.  In lowest order, the
subprocesses $gg\rightarrow \gamma c \bar{c}$ and $q\bar{q}\rightarrow
\gamma c \bar{c}$ contribute in the massive case, whereas
$c g \rightarrow \gamma c$ plus fragmentation processes contribute in the
massless case.  In a forthcoming paper, we
intend to examine the massive case in detail and to discuss comparisons with
the massless case in the regions of phase space of their respective
applicability. As remarked above, our massless approach should be appropriate
and applicable in the domain in which there is effectively only one large
scale,
$p^c_{T} \gg m_c$.

For the interval in $p^\gamma_{T}$ of current experimental interest,
10 GeV $< p^\gamma_{T} < 50$ GeV, the $g c$  and $g g$ subprocesses of
Eqs.~(\ref{eq:11}) and (\ref{eq:12}) are the most important quantitatively
at Fermilab Tevatron
energies, owing to the strength of the gluon density.  For $p^\gamma_{T} > 70$
GeV, calculations of the inclusive yield of single photons indicate that the
$q \bar{q}$ subprocess begins to dominate, but the cross section is
small in this region.  Dominance of the perturbative
subprocess initiated by $g c$ scattering is preserved after the next-to-leading
terms are included, justifying use of data from
$p +\bar{p}\rightarrow \gamma + c + X$ in attempts to measure the charm
quark momentum density in the nucleon.  However, we show that other
subprocesses account for about
$50\%$ of the cross section at currently accessible values of
$p^\gamma_{T}$.  The ``background" associated with these
subprocesses must be taken into account in analyses done to extract
the charm density.

Our results are provided in terms of the
momentum of the charm quark.  In a typical experiment,\cite{cdf} the momentum
of the
quark may be inferred from the momentum of prompt lepton decay products or
the momentum of charm mesons, such as $D^*$'s.  Alternatively, our
distributions in $z$ or $p^c_{T}$ may be convoluted with charm quark
fragmentation functions, deduced from, e.g., $e^+e^-$ annihilation
data, to provide distributions for the prompt leptons or $D^*$'s.

In Sec. II, we present our analysis of the leading and next-to-leading order
contributions to the partonic hard-scattering cross sections. Numerical results
are described in Sec. III, and a summary of our conclusions is provided in
Sec. IV.
An Appendix is included in which we present our method for performing the
required three-particle final-state integrals in n-dimensions to extract the
singularities of the two-particle inclusive hard cross section.

\section{Analytical Calculation}

We consider the two particle inclusive reaction
$A +B \rightarrow \gamma + c + X$ where $A$ and $B$ denote incident hadrons;
$p^\gamma$ and $p^c$ denote the four-vector momenta of the photon and charm
quark.  The usual Mandelstam invariants are defined in terms of the momenta of
the two incoming hadrons $P_A$ and $P_B$, and the momentum fractions of
the initial partons, $x_1$ and $x_2$, via
\begin{eqnarray}
\hat{s}&=&(x_1 P_A+x_2 P_B)^2=x_1 x_2 s \nonumber \\
\hat{t}&=&(x_1 P_A - p^{\gamma})^2 \nonumber \\
\hat{u}&=&(x_2 P_B - p^{\gamma})^2.   \label{eq:maldel}
\end{eqnarray}
Here $\sqrt{s}$ is the center-of-mass energy in the hadronic system.
We define
\begin{eqnarray}
v&=&1+\frac{\hat{t}}{\hat{s}} \nonumber \\
w&=&\frac{-\hat{u}}{\hat{s}+\hat{t}}. \label{eq:vw}
\end{eqnarray}

\subsection{Leading Order Contributions}

In leading order in perturbative QCD, only one direct subprocess
contributes to the hard-scattering cross section, the QCD Compton process
$c g\rightarrow \gamma c$,
unlike the case for single inclusive prompt photon production, where the
annihilation process $q\bar{q}\rightarrow \gamma g$ also contributes.
Since the leading order direct partonic subprocess has a two-body final state,
the photon and $c$-quark are produced with balancing transverse
momenta, and the variable $z$, defined in Eq.~(\ref{eq:zdef}), is always unity.

The leading order direct partonic cross section is
\begin{equation}
\frac{d\hat{\sigma}}{dvdzdw}=\frac{d\hat{\sigma}}{dv}\delta(1-z)\delta(1-w),
\label{eq:borna}
\end{equation}
where $d\hat{\sigma}/dv$ is the partonic Born cross section:
\begin{equation}
\frac{d\hat{\sigma}}{dv}(c g\rightarrow \gamma c)=\frac{1}{N_C}\frac{\pi
\alpha_{em}\alpha_s e_q^2}{\hat{s}}\frac{1+(1-v)^2}{1-v}.\label{bornb}
\end{equation}
Here $\alpha_{em}$ and $\alpha_s$ are the electromagnetic and strong
coupling constants, respectively, $N_C=3$ is the number of colors, and
$e_q$ denotes the quark charge.

The full expression for the physical cross section in leading order is
\begin{equation}
\frac{d\sigma}{dp_T^{\gamma}dy^{\gamma}dz}=2\pi p_T^{\gamma}\frac{1}{\pi
s}\int^1_{V W} \frac{dv}{1-v}
f_g^A(x_1,M^2)f_c^B(x_2,M^2)\frac{d\hat{\sigma}}{dv}\delta(1-z)\delta(1-w)
+ (c \leftrightarrow g).  \label{eq:losigma}
\end{equation}
Quantities $V$ and $W$ are defined similarly to $v$ and $w$,
Eq.~(\ref{eq:vw}), but
in the hadronic system; $f^A(x_1,M^2)$ denotes the parton density in hadron $A$
as a function of the momentum fraction $x_1$ and factorization scale $M$.

In addition to the lowest order direct subprocess just discussed,
$c g\rightarrow \gamma c$, there are fragmentation contributions that are also
effectively of leading order in $\alpha_s$.  In these contributions the photon
is produced through fragmentation of a final-state parton from any of the
$O(\alpha_s^2)$ subprocesses listed below.  The fragmentation functions are
essentially of order $O(\alpha_{em}/\alpha_s)$
\begin{eqnarray}
c+g&\rightarrow& g+c \nonumber \\
g+g&\rightarrow& c+\bar{c} \nonumber \\
c+q&\rightarrow& c+q \nonumber \\
c+\bar{q}&\rightarrow&c+\bar{q} \nonumber \\
c+c&\rightarrow& c+c \nonumber \\
c+\bar{c}&\rightarrow & c+\bar{c} \nonumber \\
q+\bar{q}&\rightarrow & c+\bar{c}. \label{eq:fragproc}
\end{eqnarray}
We are interested in configurations in which the photon and charm quark
have relatively large and to-some-extent balancing values of transverse
momentum.  Therefore, in the cases of the first, third, and fourth of the
subprocesses listed above, the photon is produced from fragmentation of the
$g$ and non-charm quark $q$, respectively.  In the
other cases it is produced in the fragmentation of one of the
(anti)charm quarks. The expression we use to evaluate the fragmentation
contributions is
\begin{eqnarray}
\frac{d\sigma}{dp_T^{\gamma}dy^{\gamma}dz}&=&2\pi p_T^{\gamma}\frac{1}{\pi
s}\int^1_{1-V+V W}\frac{dz'}{z'^2}\int^1_{V W} \frac{dv}{1-v}
f_a^A(x_1,M^2)f_b^B(x_2,M^2)\frac{d\hat{\sigma}^{ab\rightarrow i X}}{dv}
\nonumber \\
& &\times D_{\gamma/i}(z',Q^2)\delta(\frac{1}{z}-z').  \label{eq:fragsigma}
\end{eqnarray}

In a fully consistent next-to-leading calculation, one should calculate
the subprocesses in Eq.~(\ref{eq:fragproc}) to $O(\alpha_s^3)$, since the
photon
fragmentation functions that are convoluted with the hard subprocess
cross sections are of $O(\alpha_{em}/\alpha_s)$.  For simplicity, we
include them in $O(\alpha_s^2)$ only.  In fact, next-to-leading order
fragmentation contributions to single prompt photon production have been
included only once before\cite{gorvogel}. We expect the next-to-leading order
corrections to the fragmentation contributions to be insignificant numerically
especially after isolation cuts are imposed.

\subsection{Next-to-leading order Contributions}

There are two classes of contributions in next-to-leading order. First there
are the
virtual gluon exchange corrections to the lowest order process. Examples are
shown in Fig.1b. These amplitudes interfere with the Born amplitudes and
contribute at $O(\alpha_{em}\alpha_s^2)$. They have been calculated
twice before.\cite{gamnlo1,gorvogel} We use the results of
Ref.~\cite{gorvogel}.  The virtual contributions are proportional to
$\delta(1-z)$ and $\delta(1-w)$. At next-to-leading order there are also
three-body final-state contributions, listed in Eq.~(\ref{eq:1}). The
matrix elements for these are also taken
from Ref.~\cite{gorvogel}, where they are calculated for
single inclusive prompt photon production.

The main task of our calculation is to integrate the three-body matrix elements
over the phase space of
the unobserved particle in the final state. The situation here is
different from the standard case of single inclusive particle
production, first developed in Ref.~\cite{rke}, since we wish to
retain as much control as possible over the kinematic variables of a
second particle in the final state, while at the same time integrating
over enough of the phase space to ensure cancellation of all infrared
and collinear divergences, inherent when massless
particles are assumed. Because our goal is to provide a
fully analytic calculation, we find it necessary to integrate over the
full range of rapidity of one of the observed final-state particles. We
choose to integrate over that of the charm quark, since the photon is
usually considered the trigger particle in the experiments.

The situation here is similar to that met by
Aurenche {\it{et al}}\cite{aur2,aur3}, and
we use a similar technique to perform the phase space integrals. We
give a fairly detailed outline of the method since it is necessary to
adapt it to our situation and also because it has not been widely used.
We believe our presentation clarifies certain details which are not
stressed in the above references.

The three-body phase space integration is
done in the rest frame of the observed $c$ (or $\bar{c}$-quark) and the third
unobserved parton. Denoting the momenta of the process by
$p_1+p_2\rightarrow k_1+k_2+k_3$, we work in the rest frame of $k_2$ and
$k_3$, where $k_1$ is the momentum of the trigger photon. The final form
of the three-particle phase space integral (see the Appendix) is
\begin{eqnarray}
PS^{(3)}&=&\frac{\pi\hat{s}}{8(2\pi)^5}\left(
\frac{4\pi}{\hat{s}}\right)^{\epsilon}
\frac{v}{\Gamma(1-2\epsilon)}\left(\frac{4\pi}{\hat{s}wv(1-v)}\right)^{\epsilon}
v^{-\epsilon}(1-w)^{-\epsilon}2\sqrt{\frac{w(1-v)}{(1-v w)}}\nonumber \\
& &\times\left[ \frac{1-w+4w(1-v)z(1-z)}{1-v w}\right]^{-\epsilon}\int^\pi_0
d\theta_2 \sin^{-2\epsilon}(\theta_2) . \label{eq:phasespace}
\end{eqnarray}
We are left to perform the final integration of the squared matrix
elements over $\theta_2$.

As in the case of single inclusive cross section calculations,
documented extensively elsewhere, one can use relations
among the Mandelstam variables to reduce complex combinations of them
to simple products and ratios. The phase space integral over $\theta_2$ is
performed in $4-2\epsilon$ dimensions, thereby exposing collinear
and soft singularities as poles in $\epsilon$.
After the three-particle phase space integrals are performed, we obtain
a three-body final state
hard-scattering cross section that we represent by the expression
\begin{displaymath}
\frac{d\sigma^R_{ij}}{dv dw dz}\left( \hat{s},v,w,z,\frac{1}{\epsilon^2},
\frac{1}{\epsilon}\right) .
\end{displaymath}
Superscript $R$ indicates that this is the subprocess cross section for a
real three-body final-state contribution, as distinct from the contribution
from the virtual gluon exchange contributions that we denote $\sigma^V_{ij}$.
The subscripts ${ij}$ designate one of the processes in Eq.~(\ref{eq:1}).  In
general,
$\sigma^R_{ij}$ has single and double poles in $\epsilon$.  In accord with
the factorization theorem of perturbative QCD, the double and some of the
single poles cancel between the real and virtual contributions.  The remaining
single poles in $\epsilon$ represent collinear divergences that are
subtracted into parton densities and fragmentation functions.

In order to illustrate how the collinear singularities are handled we discuss
a few representative examples.

(a) $c + g\rightarrow\gamma + c + X$

This is the QCD Compton process plus higher order corrections. We label
the momenta by
\begin{equation}
c(p_1) + g(p_2)\rightarrow \gamma(k_1) + c(k_2) + g(k_3) .\label{eq:c}
\end{equation}
In performing the phase space integration, we expect to encounter
singularities where the gluon $k_3$ becomes soft and/or parallel to
$p_1,p_2$ or $k_2$. Since we require that the observed charm quark and
$\gamma$ be in opposite hemispheres, we will not encounter any
singularity where $k_1$ and $k_2$ are collinear (see the Appendix). In the
cases where the gluon is either soft and/or parallel to $p_1$ or $p_2$,
then $z=1$. We expose the $z\rightarrow 1$ singularities by using the
expansion
\begin{equation}
\frac{1}{|1-z|^{1+2\epsilon}}=\frac{1}{-2\epsilon}\delta(1-z)+\frac{\theta(1-z)
}{(1-z)_+}+\frac{\theta(z-1)}{(z-1)_+}-2\epsilon
\left(\frac{\ln(1-z)}{1-z)}\right)_+\theta(1-z)+O(\epsilon^2).\label{eq:zexp}
\end{equation}
There are plus-distributions in the variable $z$, as well as the usual
ones in $w$ that arise in the single particle inclusive case
and correspond to the gluon becoming either soft or collinear to $k_2$.
Plus-distributions in $z$ and $w$ can be encountered simultaneously and
must be treated carefully in the numerical evaluation of the cross
section.

Once the phase space integrals are performed and the soft and
collinear poles are exposed, we can add the real three-body
contributions to the virtual gluon exchange terms, after which all the double
poles cancel along with some single poles. The remaining collinear poles must
be factored into the parton distribution and fragmentation
functions. We perform these subtractions in the universal or
$\overline{MS}$ scheme, described in detail in many places.

To account for all collinear configurations
allowed in the subprocess, the counter cross section or factorization
formula that must be added to our results in order to cancel the
collinear poles is
\begin{eqnarray}
\frac{1}{\hat{s}v}\frac{d\sigma^F}{dv dw dz}&=&
-\frac{\alpha_s}{2 \pi}\left[\frac{1}{\hat{s}v}H_{cc}(w,M^2)
\frac{d\sigma^{cg\rightarrow\gamma
c}}{dv}(w\hat{s},v,\epsilon)\delta(1-z) \right. \nonumber \\
&+&\left. \frac{1}{\hat{s}(1-v w)}H_{gg}\left( \frac{1-v}{1-v w},M^2\right)
\frac{d\sigma^{cg\rightarrow\gamma c}}{dv}(w\hat{s},v
w,\epsilon)\delta(1-z)\right. \nonumber \\
&+&
\left.
\frac{1}{\hat{s}v}\tilde{H}_{cc}(z,M''^2)\frac{d\sigma^{cg\rightarrow\gamma c}}
{dv}(\hat{s},v,\epsilon)\theta(1-z)\delta(1-w) \right] .\label{eq:faccg}
\end{eqnarray}

\begin{equation}
H_{ij}(z,Q^2)=-\frac{1}{\hat{\epsilon}}P_{ij}(z)
\left[\frac{\mu^2}{Q^2}\right]^\epsilon+f_{ij}(z) ,\label{eq:hfunc}
\end{equation}
and
\begin{equation}
\tilde{H}_{ij}(z,Q^2)=-\frac{1}{\hat{\epsilon}}P_{ij}(z)
\left[\frac{\mu^2}{Q^2}\right]^\epsilon+d_{ij}(z) .\label{eq:hfunc2}
\end{equation}
Functions $P_{ij}(z)$ are the one-loop splitting
functions [18], $f_{ij}(z)=0$ and $d_{ij}(z)=0$ in the
$\overline{MS}$ factorization
scheme, and $\mu$ is the renormalization scale.
In the $\overline{MS}$ scheme,
$1/\hat{\epsilon}\equiv1/\epsilon-\gamma_E+\ln4\pi$.

In Eq.~(\ref{eq:faccg}), we distinguish
the factorization scale $M$ and the quark to quark plus gluon fragmentation
scale $M''$.  The last term indicates that we factor the collinear
singularity that arises when the observed charm quark $k_2$ becomes parallel
to the gluon, $k_3$, into a fragmentation function at scale $M''^2$, for the
production of a charm quark. Note that this singularity occurs in the
region $z\leq 1$, since the photon must balance the momentum of the
charm-gluon system.

We are free to convolute our cross section with a fragmentation function
that describes the formation of specific charm decay
products (e.g., D or D$^*$ mesons), but we choose not to do so in this
paper.

(b) $g + g\rightarrow \gamma + c + \bar{c}$

In the gluon-gluon fusion process, $g g\rightarrow \gamma c \bar{c}$,
the photon may become collinear to the unobserved final-state quark,
a situation not encountered in the $g c$ process discussed above. This
singularity occurs at $z=z_1$ where $z_1=1/(1-v+v w)$, and, as discussed
in the Appendix, we use an expansion similar to that in
Eq.~(\ref{eq:zexp}) to expose
the singularity. Note that this singularity occurs in the region $z\geq 1$,
and that $z$ is exactly the reciprocal of the usual fragmentation
variable for a parton to fragment into a particle with a fraction
of its momentum, $1/z$. The factorization formula for this process is
\begin{eqnarray}
\frac{1}{\hat{s}v}\frac{d\sigma^F}{dv dw dz}&=&
-\frac{\alpha_s}{2 \pi}\left[\frac{1}{\hat{s}v}H_{cg}(w,M^2)
\frac{d\sigma^{cg\rightarrow\gamma
c}}{dv}(w\hat{s},v,\epsilon)\delta(1-z)\right. \nonumber \\
&+&\left. \frac{1}{\hat{s}(1-v w)}H_{cg}\left( \frac{1-v}{1-v w},M^2\right)
\frac{d\sigma^{gc\rightarrow\gamma c}}{dv}(w\hat{s},v
w,\epsilon)\delta(1-z)  \right.  \\ \label{eq:facgg}
&+&\left. \frac{1}{\hat{s}(1-v+v w)}\tilde{H}_{\gamma
\bar{c}}(1-v+v w,M'^2)\frac{d\sigma^
{g g\rightarrow c \bar{c}}}{dv}(\hat{s},\frac{v w}{1-v+v
w},\epsilon)\delta(z_1-z) \right].\nonumber
\end{eqnarray}
In this equation, we distinguish
the factorization scale $M$ and the quark to photon fragmentation
scale $M'$.

(c) $q + \bar{q}\rightarrow \gamma + c + \bar{c}$

The process $q \bar{q}\rightarrow \gamma c \bar{c}$, as well as that
of Eq.~(\ref{eq:16}),
has a final-state collinear singularity when a gluon splits into a
collinear $c\bar{c}$ pair, and, in addition,  a singularity when the photon is
produced from fragmentation of a final-state quark.
The factorization formula for this case is
\begin{eqnarray}
\frac{1}{\hat{s}v}\frac{d\sigma^F}{dv dw dz}&=&
-\frac{\alpha_s}{2
\pi}\left[\frac{1}{\hat{s}v}\tilde{H}_{cg}(z,M''^2)\frac{d\sigma^
{q \bar{q}\rightarrow \gamma
g}}{dv}(\hat{s},v,\epsilon)\theta(1-z)\right. \nonumber \\
&+&\left. \frac{1}{\hat{s}(1-v+v w)}\tilde{H}_{\gamma \bar{c}}(1-v+v
w,M'^2)\frac{d\sigma^
{q \bar{q}\rightarrow c \bar{c}}}{dv}(\hat{s},\frac{v w}{1-v+v
w},\epsilon)\right. \nonumber \\
& &\times\left. \delta(z_1-z)\right]. \label{eq:facqqb}
\end{eqnarray}

\subsection{Physical cross section}

Once all singularities are dealt with, we calculate the physical
cross section by convoluting the hard partonic cross section with parton
distribution functions. In terms of the variables we are using, the
cross section at next-to-leading order is
\begin{eqnarray}
\frac{d\sigma}{dp_T^{\gamma}dy^{\gamma}dz}&=&2\pi p_T^{\gamma}\frac{1}{\pi s}
\sum_{i,j}\int^1_{VW}\frac{dv}{1-v}\int^1_{V
W/v}\frac{dw}{w}f_i^A(x_1,M^2)f_j^B(x_2,M^2) \nonumber \\
& &\left[
\frac{1}{v}\frac{d\hat{\sigma}^{ij}}{dv}\delta(1-z)\delta(1-w)+
\frac{\alpha_s(\mu^2)}{2\pi}K_{ij}(\hat{s},v,w,z,\mu^2,M^2,M'^2,M''^2)\right].
\label{eq:nlosigma}
\end{eqnarray}
The first term within the square brackets is the leading order part,
and
\begin{displaymath}
K_{ij}(\hat{s},v,w,z,\mu^2,M^2,M'^2,M''^2)
\end{displaymath}
is the next-to-leading order
correction term; $K_{ij}$ may include virtual gluon exchange contributions.

Taking the $c g$ subprocess as an example, we outline
how we obtain the function $K_{ij}(\hat{s},v,w,z,\mu^2,M^2,M'^2,M''^2)$.
The virtual gluon exchange contributions are represented by
\begin{displaymath}
\frac{d\sigma^V_{cg}}{dvdwdz}\left( \hat{s},v,\mu^2,
\frac{1}{\epsilon^2},\frac{1}{\epsilon} \right).
\end{displaymath}
They are proportional to $\delta(1-w)$ and $\delta(1-z)$.  The real three-body
contributions are denoted
\begin{displaymath}
\frac{d\sigma^R_{cg}}{dv dw dz}\left( \hat{s},v,w,z,\frac{1}{\epsilon^2},
\frac{1}{\epsilon}\right) .
\end{displaymath}

Combining the three-body final-state contribution and the virtual gluon
exchange
contribution and adding to these the subtraction term in
Eq.~(\ref{eq:faccg}), we derive
a finite subprocess cross section:
\begin{eqnarray}
K_{cg}(\hat{s},v,w,z,\mu^2,M^2,M''^2)&=&\frac{d\sigma^V_{cg}}{dvdwdz}
\left( \hat{s},v,\mu^2,\frac{1}{\epsilon^2},\frac{1}{\epsilon}\right)+
\nonumber \\
& &\frac{d\sigma^R_{cg}}{dv dw dz}\left( \hat{s},v,w,z,\frac{1}{\epsilon^2},
\frac{1}{\epsilon}\right)+ \nonumber \\
& &\frac{d\sigma^F_{cg}}{dvdwdz}\left( \hat{s},v,w,z,
\frac{1}{\epsilon},M^2,M''^2\right).\label{eq:f}
\end{eqnarray}
At this stage all single and double poles cancel, and we are left with a
finite cross section dependent on the factorization scale $M$ and
fragmentation scale $M''$. Because of the additional variable $z$, the
function $K_{cg}$ is quite lengthy when compared
to that for inclusive single photon production.\cite{gorvogel}
In schematic notation, where only the
$z$-distributions are made explicit, we can write the hard-scattering cross
section as
\begin{eqnarray}
K_{cg}(\hat{s},v,w,z,\mu^2,M^2,M''^2)&=&c_1(v,w)\delta(1-z)+c_2(v,w)
\frac{\theta(1-z)}{(1-z)_+}\nonumber \\
&+&c_3(v,w)\frac{\theta(z-1)}{(z-1)_+}+c_4(v,w)
\left(\frac{\ln(1-z)}{1-z}\right)_+\nonumber \\
&+&c_5(v,w,z).\label{eq:g}
\end{eqnarray}
The functions $c_i(v,w)$ contain, in general, distributions in
$(1-w)$, and they
can be expressed by
\begin{eqnarray}
c_i(v,w)&=&c_i^1(v)\delta(1-w)+c_i^2(v)\frac{1}{(1-w)_+}+c_i^3(v)\left(
\frac{\ln(1-w)}{1-w}\right)_+\nonumber \\
&+&c_i^4(v,w) .\label{h}
\end{eqnarray}
Similar expressions can be written for the other subprocesses.
These will generally involve the fragmentation scale on the photon leg,
$M'$, and additional distributions in $(z_1-z)$ and $(z-z_1)$. These are
defined as normal plus-distributions, but in the intervals $[0,z_1]$, and
$[z_1,z_{max}]$ , respectively. We integrate the
distributions between limits other than these.  For example, if the
limits in the first case are $[z_a,z_1]$, we must make the
replacement
\begin{equation}
\frac{1}{(z_1-z)_+}=\frac{1}{(z_1-z)_{z_a}}+\delta(z_1-z)\ln(z_1-z_a) ,
\label{i}
\end{equation}
where the new distribution is defined by
\begin{equation}
\int^{z_1}_{z_a}dz\frac{f(z)}{(z_1-z)_{z_a}}=
\int^{z_1}_{z_a}dz\frac{f(z)-f(z_1)}{z_1-z}.\label{j}
\end{equation}

By expanding our integrated
matrix elements as plus-distributions in $z$, we are able to expose the
singularities that occur at $z=1$ and $z=z_1$. This procedure ensures that
these integrable singularities can be treated numerically.  However, it also
means that our analytic distributions in z are singular at $z=1$
and $z=z_1$.  For comparison with experiment, we provide predictions for the
$z$ dependence in the form of histograms with finite bin-widths $\Delta z$,
reminiscent of experimental resolution.  As in Ref.~\cite{aur2}, we define
\begin{equation}
\frac{d\sigma}{dp_T^{\gamma}dy^{\gamma}dz}=\frac{1}{\Delta z}
\int^{z+\frac{\Delta
z}{2}}_{z-\frac{\Delta z}{2}}
\frac{d\sigma}{dp_T^{\gamma}dy^{\gamma}dz'}dz' .\label{k}
\end{equation}
For distributions in $p_T^{\gamma}$, we integrate over a specified range
of $z$,
\begin{equation}
\frac{d\sigma}{dp_T^{\gamma}dy^{\gamma}}=
\int^{z_b}_{z_a}
\frac{d\sigma}{dp_T^{\gamma}dy^{\gamma}dz}dz .\label{l}
\end{equation}

This completes our discussion of the calculation. Further details can be
found in the Appendix.

\section{Numerical Results and Discussion}

In this section we present and discuss explicit
evaluations of the correlated production cross section
of charm plus a prompt photon.
We provide results at $\bar p p$ center-of-mass energy
$\sqrt{s} = $ 1.8 TeV appropriate for the CDF and D0
experimental investigations underway at Fermilab.
The cross sections we evaluate are those derived in the text:
Eqs.~(\ref{eq:losigma}), (\ref{eq:fragsigma}) and (\ref{eq:nlosigma}). For
the electromagnetic coupling strength we use $\alpha_{em} = 1/137$,
and we employ a two-loop expression for
$\alpha_s (\mu^2)$ with quark threshold effects handled properly.
We choose identical values for the renormalization, factorization, and
fragmentation scales, $\mu=M=M'=M''$.  In the results presented below,
we vary $\mu$ to examine the sensitivity of the cross section
to its choice.
We choose $\Lambda^{(4)}_{QCD}$ according to the parton distribution set
we use; $\Lambda^{(4)}_{QCD}=0.200$ for the GRV parton
distributions.\cite{parden1}  The sums run over 4 flavors of quarks
$(u, d, c, s)$, all assumed massless.  We do not
include a $b$ quark contribution in our calculation.

Most of the calculations reported here are done with the
GRV parton densities\cite{parden1}.  We observe some differences
when we use instead the CTEQ3M densities\cite{parden2}.  The magnitude
and Bjorken $x$ dependence of the charm quark density in these two sets
are similar, as shown in Fig. 2, but show some differences at large $x$,
leading to a $30\%$ difference in the cross section at
$p_T^{\gamma}=60 GeV$.  In these densities, the charm
quark probability is generated through perturbative evolution, and there
is no non-perturbative intrinsic charm\cite{STANB} component.  Neither
density may be correct since there
is little direct experimental information to constrain this
density\cite{ELBW}.
A goal of our analysis is to ascertain the extent to which the $gc$ initial
state is expected to dominate the cross section for
$p +\bar{p}\rightarrow \gamma + c + X$, and, thus, the extent to which data
from this reaction may serve to measure the charm quark density.

The quark-to-photon fragmentation function is expressed as
\begin{eqnarray}
z\, D_{q \rightarrow \gamma} (z,\mu^2) &=&
 \frac{\alpha _{em}}{2\pi} \left[
  e_{q}^{2}\
  \frac{2.21-1.28z+1.29z^{2}}{1-1.63\,\ell n\left(1-z\right)}\,
  z^{0.049} +0.002 \left( 1-z \right) ^{2} z^{-1.54}
 \right] \nonumber \\
&&\times \ell n \left( \mu^{2} / \mu^{2}_0 \right).
\label{eq:fragfunc}
\end{eqnarray}
The gluon-to-photon fragmentation function is
\begin{equation}
z\, D_{g \rightarrow \gamma} (z,\mu^2)
= \frac{\alpha _{em}}{2\pi}\,
  0.0243 \left( 1-z \right)
  z^{-0.97}\, \ell n \left( \mu^{2} / \mu^{2}_0 \right).
\label{vvv}
\end{equation}
These expressions for $D_{q\rightarrow \gamma}$ and
$D_{g\rightarrow \gamma}$, taken from Ref.~\cite{frag1}, are used as
a guideline for our estimates. The physical significance of scale
$\mu_0$ is that the fragmentation function vanishes for energies less
than $\mu_0$.  For the $u, d, s$, and $c$ quarks, we set
$\mu_0 = \Lambda^{(4)}_{QCD}$, as in Ref.~\cite{frag1}.
We remark that we use simple leading order fragmentation
functions in our calculation, in contrast to the fact that we have done
a next-to-leading order $\overline{\mbox{MS}}$ calculation.  It would
be more consistent and, therefore, preferable to use
$\overline{\mbox{MS}}$ fragmentation functions evolved in
next-to-leading order.  Our choice of leading-order fragmentation
functions is motivated by our desire to work with
analytic expressions.  In published analyses of next-to-leading order
fragmentation functions,\cite{frag2,frag3} the general formalism is
presented but the fragmentation functions themselves must be obtained
through numerical evolution codes. Our
primary purpose in this paper is to provide a theoretical framework
for the analysis of the correlated production of charm and prompt photon
not necessarily to present the most up-to-date
numerical predictions. Thus, we believe our leading-order fragmentation
functions are adequate.

In several figures to follow, we show the predicted behavior of the
photon yield as a function of $p^\gamma_{T}$ and $z$,
as well as the breakdown of the total yield into contributions
from the leading order and the various next-to-leading order pieces.
The ratio $z$ is defined in Eq.~(\ref{eq:zdef}).
We choose to display cross sections as a function of
the ratio $z$, for fixed values of $p^\gamma_{T}$, or as a function
of $p^\gamma_{T}$.  We choose the
renormalization/fragmentation scale
$\mu = p^\gamma_{T}$.  Since both the photon and final charm particle
carry large transverse momentum, we could perhaps equally well choose
$\mu = p^c_{T}$ or some combination of the two.  In selecting $p^\gamma_{T}$,
we focus upon the photon as the ``trigger" particle whose transverse momentum
is well determined.  We display the $\mu$ dependence of our results below.

Throughout this paper, for clarity and simplicity of the discussion,
we refer consistently to charm production, e.g.,
$p +\bar{p}\rightarrow \gamma + c + X$.  However, the numerical values of
the cross sections shown in the figures are those for the sum of
charm and anticharm production in $p \bar{p}$ scattering.  In Fig. 3, we
present the photon yield
as a function of the ratio z for two choices of $p^\gamma_{T}$.
The same results are displayed in Fig. 4 as a function of
$p^\gamma_{T}$ for $z$ integrated over the interval 0.2 to 2.0.
We restrict $z > 0.2$ as otherwise the transverse momentum of the
charm quark could become unacceptably small.
In Fig. 3(a), the net lowest order contribution is shown at
$p^\gamma_{T}$ = 15 GeV.  The lowest
order contribution is made up of the lowest order direct term, $c g \rightarrow
\gamma c$, and the fragmentation terms discussed in Sec. II.A.  The direct
term provides a $\delta$-function at $z=1$ since the photon and charm quark
carry equal but opposite transverse momenta at this order.  The parton to
photon fragmentation contributions populate the region $z > 1$.  In the
collinear fragmentation, the photon's transverse momentum is opposite
to that of the charm quark but its magnitude is less. One of the
striking features of Fig.3a, is that the net fragmentation contribution
to the cross section is quite small compared to the case of inclusive
photon production.  At Tevatron energies, fragmentation accounts for about
$50\%$ of the inclusive yield at this value of $p_T$ \cite{field,gorvogel}.
(Note that we have not yet imposed any isolation
cuts on the cross section.) One reason for the small fragmentation contribution
is that fragmentation from the $cg$ initiated process is strongly
suppressed due to our restriction that the charm quark and photon be in
opposite hemispheres ($z \geq 0$). Thus only fragmentation from the gluon
leg is included, and the $g\rightarrow \gamma$ fragmentation function is in
general smaller than that for $q\rightarrow \gamma$.

In Figs. 3(b) and Fig. 3(c), we show the $z$ distribution after the next-to-
leading order contributions are included.  The solid lines show the full result
in which both the lowest order and all next-to-leading order terms are
incorporated.  Comparing the solid curve in Fig. 3(b) with that in Fig. 3(a),
we note that the $z$ distribution is substantially altered once the
next-to-leading order terms are included.  In particular, the peak at $z$ = 1
is reduced in magnitude by about a factor of 2, and the $z$ distribution gains
significant breadth below and above $z$ = 1.  The reduction in the magnitude of
the peak at $z = 1$ is attributed to the effect of the $O(\alpha ^2_s)$
collinear contributions on the initial parton legs. These collinear terms
provide the same event structure as the lowest order direct subprocess, viz.,
a final-state photon and charm quark with equal but opposite transverse
momenta, but their contribution is negative due to $\ln(1-z)$ terms
from the phase space and large logarithms of
$(1-z_{min})$ and $(z_{max}-1)$ from the $1/(1-z)_+$ and
$(z-1)_+$ distributions; $z_{min}$ and $z_{max}$ are the lower and
upper edges of the bins around $z=1$. On the other hand, away from collinear
configurations, the $O(\alpha_{em}\alpha ^2_s)$
subprocesses, listed in Eq.~(\ref{eq:1}), generate three body final states in
which three final partons share the transverse momentum balance.  The
non-collinear contributions therefore populate a broad interval
in $z$.

In addition to the complete result through next-to-leading order, the solid
line in Figs. 3(b) and 3(c), we display also contributions from three of the
$O(\alpha^2_s)$ terms.  The sum of the contributions from the other four
$O(\alpha^2_s)$ terms is negligible by comparison at $p^\gamma_{T}$ = 15 GeV.
The individual contributions show the important role that
the $O(\alpha^2_s)$ terms play at values of $z$ below and above 1.
Contrasting Figs. 3(b) and 3(c), we see that the peak near $z$ = 1 is predicted
to sharpen as $p^\gamma_{T}$ is increased, reflecting a diminishing importance
of
the $O(\alpha^2_s)$ terms at larger transverse momentum.

In Fig. 4, we show the cross section as a function of the transverse momentum
of the photon, $p^\gamma_{T}$.  To obtain these results, we integrate over
the interval 0.2 $< z <$ 2.0.  These results show that the $cg$ intial state
dominates the cross section until $p^\gamma_{T}$ approaches 100 GeV.  It
accounts for $60\%,\ 55\%$, and $50\%$ of total at $p^\gamma_{T}$ = 15, 45, and
60 GeV, respectively.  The $gg$
contribution is important at small values of $p^\gamma_{T}$, but it falls
off more steeply with $p^\gamma_{T}$ than the $cg$ contribution.  The
contribution from the valence subprocess,
$q \bar{q} \rightarrow c \bar{c} \gamma$, is negligible at small
$p^\gamma_{T}$,
but it overtakes the contribution of the $cg$ subprocess at sufficiently large
$p^\gamma_{T}$.  Owing to the fact the valence quarks carry significantly
harder
fractional momentum than the gluons and charm quarks, a major role for the
valence subprocess is expected at large enough $p^\gamma_{T}$.  However, the
numerical results indicate that the hard-scattering matrix element
overcomes this effect at modest values of $p^\gamma_{T}$, resulting in
dominance of the $cg$ initial state.  Comparison of Fig. (4) and Figs. 3(b)
and 3(c) shows significant $z$ variation in the fraction of the total cross
section accounted for by various subprocesses.

Dependence on the renormalization/factorization scale $\mu$ is displayed in
Figs. 5 and 6.  As $\mu$ is increased, $\alpha_s$ decreases, resulting in a
reduction of the hard-scattering cross sections.  The parton densities also
steepen as $\mu$ is increased.  Both effects contribute to the typical
decrease of the cross section at fixed large $p^\gamma_{T}$ as $\mu$ is
increased, as shown in Fig. 5.  The $\mu$ dependence of the $z$
distribution presented in Fig.6 is considerably more significant.  The
distribution becomes more sharply peaked at $z$ = 1 as $\mu$ is increased.
As shown in Fig. 3(a), the leading order direct contribution produces
a sharp peak at $z$ = 1, whereas the next-to-leading order contributions
broaden the distribution, as shown in Figs. 3(b) and 3(c).  The decrease
of $\alpha_s$ as $\mu$ increases diminishes the relative importance of the
next-to-leading order contributions.

The functional form of $D_{q\rightarrow \gamma} \left( z, \mu^2\right)$,
Eq.~(\ref{eq:fragfunc}), shows that the fragmentation contribution increases
logarithmically as  $\mu$ is increased.  If the fragmentation contributions
played a major role in the final answer, one would expect different $\mu$
dependence from that shown in Fig.~6.

In Fig. 7, we present the ``$K$-factor" as a function of $p^\gamma_{T}$.
Here $K$ is defined as the ratio of the complete answer through next-to-leading
order to the full leading order answer (including the leading order
fragmentation terms).  Our results show that for $z >$ 0.2, the inclusive
$K$ factor is about 2 for $p^\gamma_{T} >$ 15 GeV.  In the inclusive case,
no isolation requirement is imposed on the photon.  To make contact with
experiment, an isolation restriction is necessary.  Because fragmentation
contributions do not play a significant role in the associated production
of photon plus charm for $z >$ 0.2, we do not expect a great change of the
K-factor after isolation is imposed.  To estimate the impact of isolation, we
use a combination of analytic and Monte Carlo methods.\cite{moncarlo}  We
choose an isolation cone size $R =$ 0.7, and energy resolution parameter,
$\epsilon=2~GeV/p_T^{\gamma}$, as is done in the CDF experiment
\cite{cdf}.  We find that the K-factor is reduced to about 1.5, in
respectable agreement with experimental indications.\cite{cdf}

\section{Conclusions}

In summary, we have computed the contributions through $O(\alpha^2_s)$ in
perturbative QCD for inclusive associated production
of a prompt photon and a charm quark at large values of transverse momentum in
high energy hadron-hadron collisions.  The next-to-leading order terms alter
the expected distribution in the ratio of the magnitude of the transverse
momenta of the charm quark and prompt photon in an interesting and measurable
fashion.  The overall cross section increases by about a factor of two after
the next-to-leading terms are included.  Dominance of the perturbative
subprocess initiated by $gc$ scattering is preserved after the next-to-leading
terms are included, justifying use of data from
$p +\bar{p}\rightarrow \gamma + c + X$ in attempts to measure the charm
quark momentum density in the nucleon\.  However, other subprocesses are shown
to account for about
$50\%$ of the cross section at currently accessible values of
$p^\gamma_{T}$, and the ``background" associated with some of these
subprocesses, which are not initiated by charm quark scattering, such as
in Eqs.~(\ref{eq:12}) and (\ref{eq:13}) must be taken into account in
analyses done to extract the charm density.

\section{Acknowledgment}

We thank Dr. Bob Bailey and Dr. Stephen Mrenna for valuable discussions.
This work was supported
by the U.S. Department of Energy, Division of High Energy Physics,
Contract W-31-109-ENG-38.

\pagebreak

\appendix
\section{Three Body Cross Sections}
\setcounter{equation}{0}

In this Appendix we present a fairly detailed description of the techniques
for performing the 3-body phase space integrals in n dimensions.
We label the momenta for the general process by
$p_1 + p_2\rightarrow k_1 + k_2 + k_3$,
where $p_1$ and $p_2$ are the incoming partons, and $k_1$ and $k_2$
always label the observed photon and charm quark respectively.  We
integrate over the kinematic variables of $k_3$.

The calculation is performed in the rest frame of $k_2$ and $k_3$.  In this
frame of reference $\vec{k_2}+\vec{k_3}=0$ .
\begin{eqnarray}
s_{ij}&=&(k_i+k_j)^2 \nonumber \\
t_{i}&=&(p_1-p_i)^2 \nonumber \\
u_{i}&=&(p_2-p_i)^2 ,
 \label{eq:A.3}
\end{eqnarray}
where $i,j=1,2,3$, and $t_1=\hat{t}$ and $u_1=\hat{u}$, as defined in
Sec. II. In terms of the momenta, the variable $z$ is
\begin{equation}
z=-\frac{k^T_1.k^T_2}{|k^T_1|^2}=m.k_2 ,
 \label{eq:A.4}
\end{equation}
where $m$ is a vector that depends on the choice of axes. We choose our axes
in $n$ dimensions such that
\begin{equation}
m=\sqrt{\frac{\hat{s}}{{\hat{t}\hat{u}}}}\left(
\sinh\chi,0,...,0,\cosh\chi \right).
 \label{eq:A.5}
\end{equation}
The axes are fixed and cannot be changed to simplify any phase
space integrals we may encounter because, unlike the case of single
inclusive particle production, we will not integrate over the full range of
angles. The momenta of the particles can be parametrized in this frame as
\begin{eqnarray}
p_1&=&\frac{\hat{s}v}{2\sqrt{s_{23}}}(1,0,...,0,\sin\psi',\cos\psi')\nonumber
\\
p_2&=&\frac{\hat{s}(1-v w)}{2\sqrt{s_{23}}}(1,0,...,0,\sin\psi,\cos\psi)
\nonumber\\
k_1&=&\frac{\hat{s}(1-v+vw)}{2\sqrt{s_{23}}}(1,0,...,0,\sin\psi'',\cos\psi'')
\nonumber\\
k_2&=&\frac{\sqrt{s_{23}}}{2}(1,0,...,0,\sin\theta_1\cos\theta_2,\cos\theta_1)
\nonumber\\
k_3&=&\frac{\sqrt{s_{23}}}{2}(1,0,...,0,-\sin\theta_1\cos\theta_2,-\cos\theta_1).
 \label{eq:A.6}
\end{eqnarray}
Quantities $v$ and $w$ are defined in Sec. II.

{}From the definition of $m$, Eq.~(\ref{eq:A.5}), we can derive the
relationships
\begin{eqnarray}
\tanh\chi&=&\sqrt{\frac{w(1-v)}{1-v w}}\nonumber \\
\cos\psi&=&\cos\psi'=\tanh\chi \nonumber \\
\sin\psi&=&-\sin\psi'=-\sqrt{\frac{1-w}{1-v w}} \nonumber \\
\cos\psi''&=&\frac{1+v-v w}{1-v+v w}\tanh\chi \nonumber \\
\sin\psi''&=&-\frac{1-v-v w}{1-v+v w}\sqrt{\frac{1-w}{1-v w}}.
 \label{eq:A.7}
\end{eqnarray}

The constrained three-particle phase space is expressed as
\begin{eqnarray}
PS^{(3)}&=&
\int\frac{d^n k_1}{(2\pi)^{n-1}}\frac{d^nk_2}{(2\pi)^{n-1}}\frac{d^nk_3}
{(2\pi)^{n-1}}(2\pi)^n\delta^n(p_1+p_2-k_1-k_2-k_3)\nonumber \\
& &\times \delta^+(k_1^2)\delta^+(k_2^2)\delta^+(k_3^2)
\delta\left(v-1-\frac{\hat{t}}{\hat{s}}\right)\nonumber \\
& &\times \delta\left(w+\frac{\hat{u}}{\hat{s}+\hat{t}}\right)\delta(z-m.k_2) .
 \label{eq:A.8}
\end{eqnarray}
After some of the integrals are done with the aid of the $\delta$-functions,
the element of phase space reduces in $n=4-2\epsilon$ dimensions to
\begin{eqnarray}
PS^{(3)}&=&\frac{\pi\hat{s}}{8(2\pi)^5}\left(
\frac{4\pi}{\hat{s}}\right)^{\epsilon}
\frac{v}{\Gamma(1-2\epsilon)}\left(\frac{4\pi}{\hat{s}wv(1-v)}\right)^{\epsilon}
v^{-\epsilon}(1-w)^{-\epsilon}2\sqrt{\frac{w(1-v)}{(1-v w)}}\nonumber \\
& &\times \left[ \frac{1-w+4w(1-v)z(1-z)}{1-v w}\right]^{-\epsilon}\int^\pi_0
d\theta_2 \sin^{-2\epsilon}(\theta_2) .
 \label{eq:A.9}
\end{eqnarray}
In particular, we integrated over angle
$\theta_1$ using the function $\delta(z-m.k_2)$ and the
relation
\begin{equation}
z=\frac{1}{2}(1-\cos\theta_1\coth\chi) ,
 \label{eq:A.10}
\end{equation}
that can be derived from it. We are left with the task of
integrating the squared matrix-elements over angle $\theta_2$.

Using relations among the Mandelstam invariants, and partial
fractioning, we reduce functions involving $\theta_2$ to
only a few types for all subprocesses of interest.
We denote the general invariant by
\begin{equation}
T_i=T_{i0}(\alpha_i+\beta_i \cos\theta_2),
 \label{eq:A.11}
\end{equation}
where $\alpha_i$ and $\beta_i$ are functions of $\psi$, $\psi'$ and
$\psi''$, and hence of $v$ and $w$; $T_{i0}$ is also a function of the
latter (see Eqs.~(\ref{eq:A.6}) and (\ref{eq:A.7})).
The combinations we must consider are
\begin{displaymath}
\frac{1}{T_i},\;\;\;\;\frac{T_j^n}{T_i},\;\;\; {\rm and}\;\; \frac{1}{T_iT_j},
\end{displaymath}
where, $i,j,n=1,2,3$.
These, in turn, are all expressible in terms of two general integrals, but
the form of the functions $\alpha$ and $\beta$
determines the final result, such as its singularity
structure.

The two general integrals are
\begin{equation}
I_0[T_i]=I_0=\int^\pi_0\sin^{-2\epsilon}\theta_2
d\theta_2=\pi2^{2\epsilon}\frac{\Gamma[1-2\epsilon]}{\Gamma^2[1-\epsilon]} ,
 \label{eq:A.12}
\end{equation}
and
\begin{eqnarray}
I_1[T_i]&=&T_{i0}\int^\pi_0\frac{\sin^{-2\epsilon}\theta_2 d\theta_2}{T_i}=
\int^\pi_0\frac{\sin^{-2\epsilon}\theta_2 d\theta_2}
{(\alpha+\beta\cos\theta_2)}\nonumber \\
&=&\frac{\pi}{\sqrt{\alpha^2-\beta^2}}\left[\frac{4\alpha^2}
{\alpha^2-\beta^2}\right]^\epsilon\frac{\Gamma[1-2\epsilon]}
{\Gamma^2[1-\epsilon]}\; _2F_1\left(\frac{1}{2}-\epsilon,-\epsilon;1-\epsilon;
\frac{\beta^2}{\alpha^2}\right).
 \label{eq:A.13}
\end{eqnarray}

In terms of $I_0$, the following powers and combinations of propagators yield
\begin{eqnarray}
T_i^0&\Longrightarrow& I_0 ,\nonumber \\
T_i&\Longrightarrow& T_{i0}(\alpha_i I_0) ,\nonumber \\
T_i^2&\Longrightarrow& T_{i0}^2 \left(
\alpha^2_i+\frac{\beta^2_i}{2(1-\epsilon)}\right) I_0 ,\nonumber \\
T_i T_j&\Longrightarrow& T_{i0}T_{j0} \left(
\alpha_i \alpha_j+\frac{\beta_i \beta_j}{2(1-\epsilon)}\right) I_0 ,\nonumber
\\
T_i^3&\Longrightarrow& T_{i0}^3\left(
\alpha^3_i+\frac{3 \alpha_i \beta^2_i}{2(1-\epsilon)}\right) I_0.
 \label{eq:A.14}
\end{eqnarray}
In terms of $I_1$ we obtain:
\begin{eqnarray}
\frac{1}{T_i}&\Longrightarrow&\frac{1}{T_{i0}}I_1[T_i] ,\nonumber \\
\frac{T_j}{T_i}&\Longrightarrow&\frac{T_{j0}}{T_{i0}}\left(\frac
{\alpha_j\beta_i-
\alpha_i\beta_j}{\beta_i}I_1[T_i]+\pi\frac{\beta_j}{\beta_i}\right)
,\nonumber\\
\frac{T_j^2}{T_i}&\Longrightarrow&\frac{T_{j0}^2}{T_{i0}}\left(
\left[\frac{\alpha_j\beta_i-\alpha_i\beta_j}{\beta_i}\right]^2 I_1[T_i]+
\pi\left[\frac{2\alpha_j\beta_j\beta_i-\alpha_i\beta_j^2}{\beta_i^2}\right]
\right) ,\nonumber\\
\frac{T_j^3}{T_i}&\Longrightarrow&\frac{T_{j0}^3}{T_{i0}}\left(
\left[\frac{\alpha_j\beta_i-\alpha_i\beta_j}{\beta_i}\right]^3 I_1[T_i]+
\pi\left[\frac{6\alpha_j\beta_j\beta_i(\alpha_j\beta_i-\beta_j\alpha_i)
+\beta_j^3(\beta_i^2+2\alpha_i^2)}{2\beta_i^3}\right]
\right) ,\nonumber\\
\frac{1}{T_iT_j}&\Longrightarrow& \frac{1}{T_{i0}T_{j0}}\frac{1}
{\alpha_j\beta_i-\alpha_i\beta_j}(
\beta_i I_1[T_i]-\beta_j I_1[T_j]) .
 \label{eq:A.15}
\end{eqnarray}

In order to demonstrate how the different propagators are handled in the
calculation, we consider a few typical examples. We examine
single propagators first, then double propagators.

\subsection{Single Propagators}

\begin{displaymath}
(i)\;\;\; \frac{1}{u_2}
\end{displaymath}
In this case $\alpha = 1-\cos\psi\cos\theta_1$, and $\beta
= -\sin\psi\sin\theta_1$.
\begin{equation}
\alpha^2-\beta^2=(\cos\theta_1-\cos\psi)^2=4z^2\tanh^2\chi .
 \label{eq:A.16}
\end{equation}
There is a singularity at $z=0$, but the physical
condition that the photon and charm quark be in opposite hemispheres
guarantees $z > 0$. The integral is therefore finite and can be
treated in $4$-dimensions.  The result is
\begin{equation}
\int^\pi_0\frac{\sin^{-2\epsilon}\theta_2d\theta_2}{u_2}=\frac{\pi}{2
u_{20}}\frac{1}{\tanh\chi}\frac{1}{z} .
 \label{eq:A.17}
\end{equation}
The result for $t_2$ is similar.  In the evaluation of $u_2$ in terms of
angles, $u_{20}$ is the overall factor that does not depend on angles.

\begin{displaymath}
(ii)\;\;\; \frac{1}{s_{12}}
\end{displaymath}
Here $\alpha = 1-\cos\psi''\cos\theta_1$, and $\beta =
-\sin\psi''\sin\theta_1$.
\begin{equation}
\alpha^2-\beta^2=4\left( z+\frac{v(1-w)}{1-v+v w} \right)^2\tanh^2\chi .
 \label{eq:A.18}
\end{equation}
A singularity occurs only for negative $z$ when the photon
and charm quark are exactly collinear and in the same
hemisphere. We can treat this integral in $4$-dimensions for
the same physical reason as above, with the result
\begin{equation}
\int^\pi_0\frac{\sin^{-2\epsilon}\theta_2d\theta_2}{s_{12}}=\frac{\pi}{2
s_{120}}\frac{1}{\tanh\chi}\frac{1}{\left(z+\frac{v(1-w)}{1-v+v
w}\right)} .
 \label{eq:A.19}
\end{equation}

\begin{displaymath}
(iii)\;\;\; \frac{1}{u_{3}}
\end{displaymath}
In this case $\alpha = 1+\cos\psi\cos\theta_1$, and $\beta
= \sin\psi\sin\theta_1$.
\begin{equation}
\alpha^2-\beta^2=4(1-z)^2\tanh^2\chi.
 \label{eq:A.20}
\end{equation}
There is a singularity when $z=1$, corresponding to
$k_3$ and $p_2$ being collinear. This pole must be exposed and factored
into the parton distributions. The integral is
\begin{eqnarray}
\int^\pi_0\frac{\sin^{-2\epsilon}\theta_2d\theta_2}{u_{3}}&=&\frac{\pi}{2
u_{30}\tanh\chi}\frac{1}{|1-z|^{1+2\epsilon}}\left[ \frac{1+\tanh^2\chi(1-2z)}{
\tanh\chi}\right]^{2\epsilon}\frac{\Gamma(1-2\epsilon)}{\Gamma^2(1-\epsilon)}
\nonumber \\
& &\times _2F_1\left(\frac{1}{2}-\epsilon,-\epsilon;1-\epsilon;
\frac{\beta^2}{\alpha^2}\right) .
 \label{eq:A.21}
\end{eqnarray}
This integral can be reduced if we use the expansion
\begin{equation}
\frac{1}{|1-z|^{1+2\epsilon}}=\frac{1}{-2\epsilon}\delta(1-z)+\frac{\theta(1-z)
}{(1-z)_+}+\frac{\theta(z-1)}{(z-1)_+}-2\epsilon
\left(\frac{\ln(1-z)}{1-z)}\right)_+\theta(1-z)+O(\epsilon^2) ,
 \label{eq:A.22}
\end{equation}
and note that the hypergeometric function at $z=1$ reduces to
$2^{-2\epsilon}$.  We introduce a plus-distribution,
\begin{equation}
\int^{z_{max}}_1\frac{f(z)}{(z-1)_+}dz=\int^{z_{max}}_1\frac{f(z)-f(1)}{
(z-1)}dz,
 \label{eq:A.23}
\end{equation}
where $z_{max}=1/2(1+\coth\chi)$ from Eq.~(\ref{eq:A.10}).
When the phase space factor involving $z$ is included in the expansion,
the integral reduces to
\begin{equation}
\int^\pi_0\frac{\sin^{-2\epsilon}\theta_2d\theta_2}{u_{3}}=\frac{\pi}{2
u_{30}\tanh\chi}\frac{\Gamma(1-2\epsilon)}{\Gamma^2(1-\epsilon)}
\left[ \delta(1-z)\left(-\frac{1}{\epsilon}-\ln z_{max}\right) +
\frac{\theta(1-z)}{(1-z)_+}+\frac{\theta(z-1)}{(z-1)_+}\right] .
 \label{eq:A.24}
\end{equation}
The result for $1/t_3$ is similar.

\begin{displaymath}
(iv)\;\;\; \frac{1}{s_{13}}
\end{displaymath}
This propagator occurs when there is a quark $k_3$ in the
final state that may become collinear with the photon $k_1$,
such as in Eqs.~(\ref{eq:12})-(\ref{eq:17}).
Here, $\alpha = 1+\cos\psi''\cos\theta_1$, and $\beta
= \sin\psi''\sin\theta_1$.
\begin{equation}
\alpha^2-\beta^2=4\left( z-\frac{1}{1-v+v w}\right)^2\tanh^2\chi.
 \label{eq:A.25}
\end{equation}
There is a singularity at $z=1/(1-v+v w)=z_1$.
Using an expansion similar to that in Eq.~(\ref{eq:A.22}), but with $(1-z)$ and
$(z-1)$ replaced by $(z_1-z)$ and $(z-z_1)$, we cast the result in the
form
\begin{eqnarray}
\int^\pi_0\frac{\sin^{-2\epsilon}\theta_2d\theta_2}{s_{13}}&=&\frac{\pi}{2
s_{130}\tanh\chi}\frac{\Gamma(1-2\epsilon)}{\Gamma^2(1-\epsilon)}
\left[
\delta(z-z_1)\left(-\frac{1}{\epsilon}-\ln\left(1-\frac{z_{min}}{z_1}\right)
\right)\right.  \nonumber \\
&+&\left. \frac{\theta(z_1-z)}{(z_1-z)_+}+\frac{\theta(z-z_1)}{(z-z_1)_+}
\right] .
 \label{eq:A.26}
\end{eqnarray}
Here $z_{min}=1/2(1-\coth\chi)$, and the plus-distributions are
defined in a similar way, but with limits from $0$ to $z_1$ and $z_1$ to
$z_{max}$.

In this outline, we have omitted phase
space factors present in Eq.~(\ref{eq:A.9}) and included them only when they
are important for the expansions performed.  In principle, all
phase space factors should be included and, for example,
$v^{\epsilon}$ would be expanded as
\begin{displaymath}
1-\epsilon\ln(v)+\frac{\epsilon^2}{2}\ln^2(v)
\end{displaymath}
and combined with the final results of $(i)$
to $(iv)$ above, before $\epsilon$ is set to zero. Factors such as
$2 v\tanh\chi$ have also been omitted but are included
in our final results.

\subsection{Double Propagators}

We examine a few important examples of double propagators. Some
have been calculated previously,\cite{aur2,aur3} but
there are cases not encountered in earlier
calculations that we stress here. We include all phase space factors since
most are needed in the expansions.

\begin{displaymath}
(i)\;\;\; \frac{1}{t_{3}u_3}
\end{displaymath}
This double pole propagator was encountered in Ref.~\cite{aur2}.  The
result is
\begin{eqnarray}
\int^\pi_0\frac{\sin^{-2\epsilon}\theta_2d\theta_2}{t_3u_{3}}&=&
\frac{\pi v^{1-\epsilon}}{t_{30}
u_{30}}\frac{\Gamma(1-2\epsilon)}{\Gamma^2(1-\epsilon)}\nonumber \\
& &\times \left[( 1-v w)\left(\frac{\theta(1-z)}{(D_1(1-z))_+}
+\frac{\theta(z-1)}{(D_1(z-1))_+}\right) \right.\nonumber \\
&+&\left. \delta(1-z)\left(-\frac{1-v w}{1-w}
\ln\left( \frac{\tanh\chi+\coth\chi}{2}\right)\right. \right. \nonumber\\
&+& \left. \left.v \left(-\frac{1}{\epsilon}-\ln(1-v w)+2 \ln(1-w)\right)
\right) \right. \nonumber \\
&-&\left. (1-v)\left( \left(\frac{1}{\epsilon}+\ln(1-v)\right)\frac{1}{(1-w)_+}
\right. \right. \nonumber \\
&-& \left. \left.
2\left(\frac{\ln(1-w)}{(1-w)}\right)_+ +\frac{1}{1-w}\ln\left( \frac{1-v
w}{1-w}\right)\right) \right. \nonumber \\
&+&\left. \delta(1-z)\delta(1-w)\left(\frac{1}{2\epsilon^2}+
\frac{1}{2\epsilon}\ln(1-v)+\frac{1}{4}\ln^2(1-v)\right)(1-v)\right],
 \label{eq:A.27}
\end{eqnarray}
where $D_1=1-w+2(1-z)w(1-v)$.

\begin{displaymath}
(ii)\;\;\; \frac{1}{t_{3}u_2}
\end{displaymath}
Applying the last result in Eq.~(\ref{eq:A.15}), we get a term $1/(1-2 z)$
multiplying the
integrals for the propagators $(\beta(u_2)I[u_2]+\beta(t_3)I[t_3])$, along
with other factors. This term is singular at $z=1/2$, but the
singularity has no physical origin and must be removed before numerical
evaluation of the cross section. The integral of $1/u_2$ yields a
term $1/z$ (Eq.~(\ref{eq:A.17})) while that of $t_3$ yields plus-distributions
in $(1-z)$ and $(z-1)$ (Eq.~(\ref{eq:A.24})). To remove the false singularity
we make the replacement
\begin{displaymath}
\frac{1}{z}\rightarrow \frac{\theta(1-z)}{z}+\frac{\theta(z-1)}{z} .
\end{displaymath}
These $\theta$ functions can be combined with the plus-distributions
from the second term in $(\beta(u_2)I[u_2]+\beta(t_3)I[t_3])$ to produce

\begin{eqnarray}
\int^\pi_0\frac{\sin^{-2\epsilon}\theta_2d\theta_2}{t_{3}u_2}&=&\frac{\pi
v^{1-\epsilon} (1-w)^{-\epsilon}}{2
t_{30}u_{20}\tanh^2\chi}\frac{\Gamma(1-2\epsilon)}{\Gamma^2(1-\epsilon)}
\left[ \delta(1-z)\left(-\frac{1}{\epsilon}-\ln z_{max}\right) +
\frac{\theta(1-z)}{z(1-z)_+}\right. \nonumber \\
&+&\left. \theta(z-1)\left(\frac{1}{(z-1)_+}+
\frac{1+2z}{(1-2 z)z}\right)\right].
 \label{eq:A.28}
\end{eqnarray}
The term $1/(1-2z)$ is harmless when multiplied by $\theta(z-1)$.

\begin{displaymath}
(iii)\;\;\; \frac{1}{t_{3}s_{23}}
\end{displaymath}

This case involves singularities when $z\rightarrow 1$ and
$w\rightarrow 1$.  It is discussed in Ref.~\cite{aur3}.
The result is
\begin{eqnarray}
\int^\pi_0\frac{\sin^{-2\epsilon}\theta_2d\theta_2}{t_{3}s_{23}}&=&\frac{\pi
v^{1-\epsilon}}{\hat{s} v t_{30}}\frac{\Gamma(1-2\epsilon)}
{\Gamma^2(1-\epsilon)}
\left[-\frac{1}{\epsilon}\delta(1-w)\left(-\frac{1}{\epsilon}\delta(1-z)+
\theta(1-z)\left( \frac{1}{(1-z)_+}\right. \right. \right. \nonumber \\
&-& \left. \left. \left. \epsilon\frac{\ln(z)}{1-z}-
\epsilon\left(
\frac{\ln(1-z)}{1-z}\right)_+\right)\right) +\frac{1}{(1-w)_+}\left(-\frac{1}
{\epsilon}
\delta(1-z)+\frac{\theta(1-z)}{(1-z)_+}\right) \right. \nonumber \\
&+&\left. \frac{1}{1-w}\left(
\frac{\theta(z-1)}{(z-1)_+}-\delta(1-z)\ln(z_{max})\right)+
\left(\frac{\ln(1-w)}{1-w}\right)_+\right] .
 \label{eq:A.29}
\end{eqnarray}
A similar result is obtained for $1/(s_{23}u_3)$ and for
$1/(s_{23}s_{13})$ except that, in the latter case, the singularities occur
at $w\rightarrow1$ and $z\rightarrow z_1$.

Finally, since $s_{23}=\hat{s} v(1-w)$, we make the point that in other cases
when the propagator $1/s_{23}$ occurs in
the denominator of the matrix elements,
it must be combined with the phase space
factor $(1-w)^{-\epsilon}$ and expanded via
\begin{equation}
(1-w)^{-1-\epsilon}=-\frac{1}{\epsilon}\delta(1-w)+\frac{1}{(1-w)_+}+\epsilon
\left( \frac{\ln(1-w)}{1-w}\right)_+ +O(\epsilon).
 \label{eq:A.30}
\end{equation}
The results of this expansion are combined with the phase space factor in
Eq.~(\ref{eq:A.9}). To ensure in every case that we retain all finite terms
and obtain the correct result in the limit when $k_2$ and $k_3$ become
collinear, we always make the full replacement
\begin{eqnarray}
\int^\pi_0\frac{\sin^{-2\epsilon}\theta_2d\theta_2}{s_{23}}&=&
\frac{2^{1+2\epsilon}\pi \tanh\chi v^{-\epsilon}}{\hat{s}}
\frac{\Gamma(1-2\epsilon)}
{\Gamma^2(1-\epsilon)}
\left[(1-z)\left(-\frac{1}{\epsilon}\delta(1-w)\left(-\frac{1}{\epsilon}
\delta(1-z)\right. \right. \right.  \nonumber \\
&+&\left. \left. \left. \theta(1-z)\left(\frac{1}{(1-z)_+}-\epsilon
\frac{\ln(z)}{1-z}-
\epsilon\left(\frac{\ln(1-z)}{1-z}\right)_+\right)\right)\right) \right.
\nonumber \\
&+&\left. \frac{1}{(1-w)_+}\left(-\frac{1}{\epsilon}\delta(1-z)+
\frac{\theta(1-z)}{
(1-z)_+}\right)+\delta(1-z)\left( \frac{\ln(1-w)}{1-w}\right)_+\right] .
 \label{eq:A.31}
\end{eqnarray}
Most of these terms will vanish since, for example, $(1-z)$ will usually
multiply $\delta(1-z)$ and $(1-w)$ will multiply $\delta(1-w)$. In a
few special cases, as when there is a $P_{qq}$ splitting function
in the collinear limit, as is the case when the final-state gluon becomes
parallel to the $c$-quark in the $cg$ initiated process, the full
expansion is needed in order to expose the singularity. This singularity
may then be factored into the $c$-quark fragmentation function.
\pagebreak

\pagebreak

\noindent
{\bf Figure Captions}
\newcounter{num}
\begin{list}%
{[\arabic{num}]}{\usecounter{num}
    \setlength{\rightmargin}{\leftmargin}}

\item (a) Lowest order Feynman diagrams for $\gamma$ plus $c$ quark
production; $k_1$ and $k_2$ are the four-vector momenta of the photon and
charm quark. (b) Examples of virtual corrections to the lowest order diagrams.
(c) Examples of next-to-leading order three-body final-state diagrams for
the $g c$ initial state.

\item Charm quark density $c(x,Q)$ as a function of Bjorken $x$ at $Q =10$ GeV.
The solid line shows the expectation
of the GRV parton densities\cite{parden1}, and the dotted line that of the
the CTEQ3M densities\cite{parden2}.

\item Cross section $d\sigma/dp^\gamma_{T} dy^\gamma dz$ as a function of $z$
for $p +\bar{p}\rightarrow \gamma + c + X$ at $\sqrt{s}=1.8$ TeV.  We set
$y^\gamma = 0$.  Results are presented in the form of a histogram in bins of
width $\Delta z=0.2$.  In (a), for $p^{\gamma}_T=15$ GeV, we show the net
contribution from the lowest order direct process $g c \rightarrow \gamma c$
and from all the leading order fragmentation processes
$p_1 p_2 \rightarrow p_3 c$ followed by the collinear fragmentation
$p_3 \rightarrow \gamma X$.  In (b) and (c),
for $p^{\gamma}_T= $15 and 45 GeV, respectively, we display the full
cross section through next-to-leading order (solid line) and contributions from
three important $O(\alpha ^2_s)$ subprocesses.

\item The transverse momentum dependence of
$d\sigma/dp^\gamma_{T} dy^\gamma dz$, for $z$ integrated over the
interval $0.2 <z <2.0$.  The upper solid line shows the sum of all subprocesses
through next-to-leading order.  The dashed line shows the sum of the
$O(\alpha_s)$ and $O(\alpha ^2_s)$ contributions from the $c g$ initial state.
The $O(\alpha ^2_s)$ contributions from the $g g$ and $c q$ initial states are
shown as dash-dot and dotted curves.  The lower solid line shows the
$O(\alpha ^2_s)$ contribution from the $\bar{q}q$ (and $\bar{c}c$ and
$c c$) initial state.

\item The renormalization/factorization scale $\mu$ dependence.  For the
sum of all contributing subprocesses,
$d\sigma/dp^\gamma_{T} dy^\gamma dz$, for $y^\gamma = 0$ and $z$ integrated
over the interval $0.2 <z <2.0$, is shown as a function of $p^\gamma_{T}$
for three values of $\mu/p^\gamma_{T}$: 0.5, 1.0, and 2.

\item The renormalization/factorization scale $\mu$ dependence of
$d\sigma/dp^\gamma_{T} dy^\gamma dz$.  Results are shown as a function of
$z$ at $p^\gamma_{T} = 20$ GeV for three values of
$\mu/p^\gamma_{T}$: 0.5, 1.0, and 2.

\item The K factor defined in the text is shown as a function of
$p^\gamma_{T}$ for inclusive (i.e., non-isolated) photons (solid
line) and isolated photons (dashed line); $y^\gamma = 0$ and
$0.2 <z <2.0$.
\end{list}

\begin{references}

\bibitem{cdf} CDF Collaboration, R. Blair {\it{et al}}, FERMILAB-Conf-95/245-E,
to be published in the Proceedings of the 10th Topical Workshop on
Proton-Antiproton Collider Physics, FNAL, May, 1995.
\bibitem{qnlo} P. Nason, S. Dawson,
and R.K. Ellis, Nucl. Phys. {\bf B303}, 607 (988); {\bf B327}, 49 (1989);
Erratum: {\bf B335}, 260 (1990); W. Beenakker, H. Kuijf, W.L. van Neerven,
and J. Smith, Phys. Rev. {\bf D40}, 54 (1989); W. Beenakker, W.L. van Neerven,
R. Meng, G.A. Schuler, and J. Smith, Nucl. Phys. {\bf B351}, 507 (1991);
E. L. Berger and R. Meng, Phys. Rev. {\bf D49} 3248, (1994), and references
therein.
\bibitem{gamnlo1} P. Aurenche {\it{et al}}, Phys Lett. {\bf 140B}, 87 (1984);
 P.~Aurenche {\it{et al}}, Nucl. Phys. {\bf B297}, 661 (1988);
 A.~P.~Contogouris, S.~Papadopoulos, and D.~Atwood, Theor. Math. Phys.
 {\bf 87}, 374 (1991);
 H.~Baer, J.~Ohnemus, and J.~F.~Owens, Phys. Rev. {\bf D42}, 61 (1990).
\bibitem{gorvogel} L.~E.~Gordon and W.~Vogelsang, Phys. Rev. {\bf D48}, 3136
(1993) and {\bf D50}, 1901 (1994);
 M.~Gl\"{u}ck, L.~E.~Gordon, E.~Reya, and W.~Vogelsang, Phys. Rev. Lett.
 {\bf 73}, 388 (1994).
\bibitem{berqiu} E. L. Berger and J.-W. Qiu, Phys. Lett. {\bf B248}, 371
(1990); and Phys. Rev. {\bf D44}, 2002 (1991).
\bibitem{aur2} P. Aurenche, A. Douiri, R. Baier, M. Fontannaz, and D.
Schiff, Z. Phys. {\bf C29}, 459 (1985).
\bibitem{boboo} B. Bailey, J.Ohnemus, and J.F.Owens, Phys. Rev. {\bf D46},
2018 (1992).
\bibitem{aur3} P. Aurenche, A. Douiri, R. Baier, M. Fontannaz, and D.
Schiff, Z. Phys. {\bf C24}, 309 (1984).
\bibitem{correl}E. L. Berger, Phys. Rev. {\bf D37}, 1810 (1988); M. Mangano,
P. Nason, and G. Ridolfi Nucl. Phys. {\bf B373}, 295 (1992).
\bibitem{sdellis} S.~D.~Ellis, Z.~Kunszt, and D.~E.~Soper, Phys. Rev. Lett.
{\bf 69}, 1496 (1992); W.~T.~Giele, E.~W.~N.~Glover, and D.~A.~Kosower,
Phys. Rev. Lett. {\bf 73}, 2019 (1994); S.~D.~Ellis and D.~E.~Soper,
Phys. Rev. Lett. {\bf 74}, 5182 (1995).
\bibitem{ELBW} E. L. Berger {\it{et al}}, Phys. Rev. {\bf D40}, 83 (1989);
U. Baur {\it{et al}}, Phys. Lett. {\bf B318}, 544 (1993).
\bibitem{field} E. L. Berger, E. Braaten, and R. D. Field, Nucl. Phys.
 {\bf B239}, 52 (1984).
\bibitem{frag1}  D.W. Duke and J. F. Owens, Phys. Rev. {\bf D26}, 1600
(1982); J. F. Owens, Rev. Mod. Phys. {\bf 59}, 465 (1987).
\bibitem{frag2} P. Aurenche {\it{et al}}, Nucl. Phys. {\bf B399}, 34 (1993).
\bibitem{frag3} M. Gl\"{u}ck, E. Reya, and A. Vogt, Phys. Rev. {\bf D48}, 116
(1993).
\bibitem{moncarlo} B. Bailey, E. L. Berger, and L. E. Gordon, Argonne report
ANL-HEP-PR-95-87, to be published.
\bibitem{strvog} M.~Stratmann and W.~Vogelsang, Phys. Rev. {\bf D52},
1535 (1995).
\bibitem{rke} R. K. Ellis, M. A. Furman, H. E. Haber, and I. Hinchliffe,
Nucl. Phys. {\bf B173}, 397 (1980).
\bibitem{altar} G. Altarelli and G. Parisi, Nucl. Phys.
{\bf B126}, 298 (1977).
\bibitem{parden1} M. Gl\"{u}ck, E. Reya, and A. Vogt, Phys. Rev. {\bf D45},
3986 (1992).
\bibitem{parden2} H. L. Lai {\it et al}, CTEQ Collaboration, Phys.
Rev. {\bf D51}, 4763 (1995).
\bibitem{STANB} S.~J.~Brodsky, P.~Hoyer, C.~Peterson, and N.~Sakai, Phys. Lett.
{\bf B93}, 451 (1980); S.~J.~Brodsky, C.~Peterson, and N.~Sakai,
Phys. Rev. {\bf D23}, 2745 (1981).
\end{references}
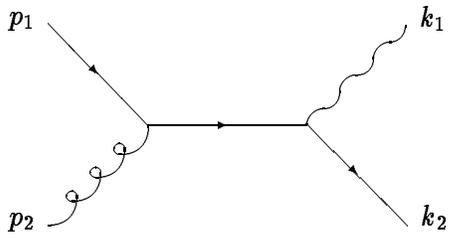
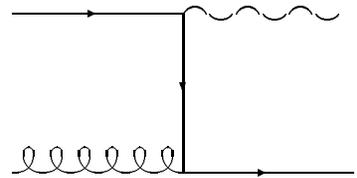
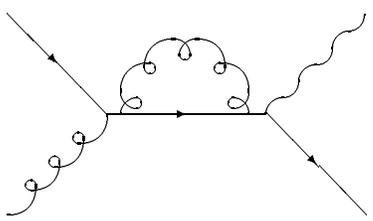
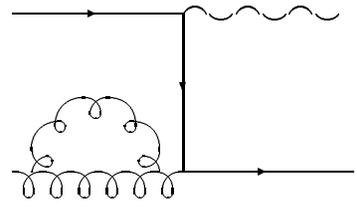
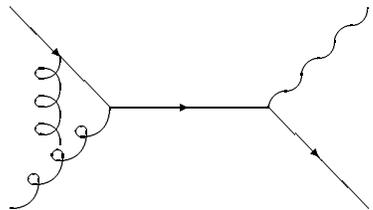
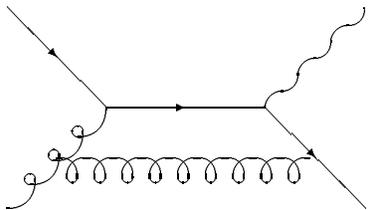
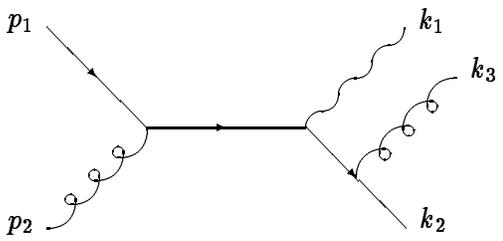
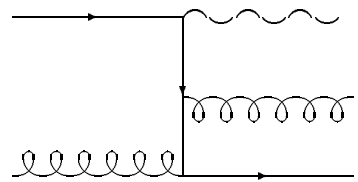
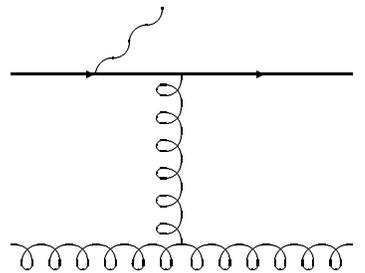
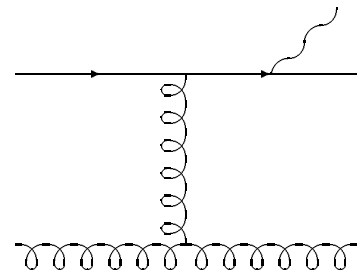
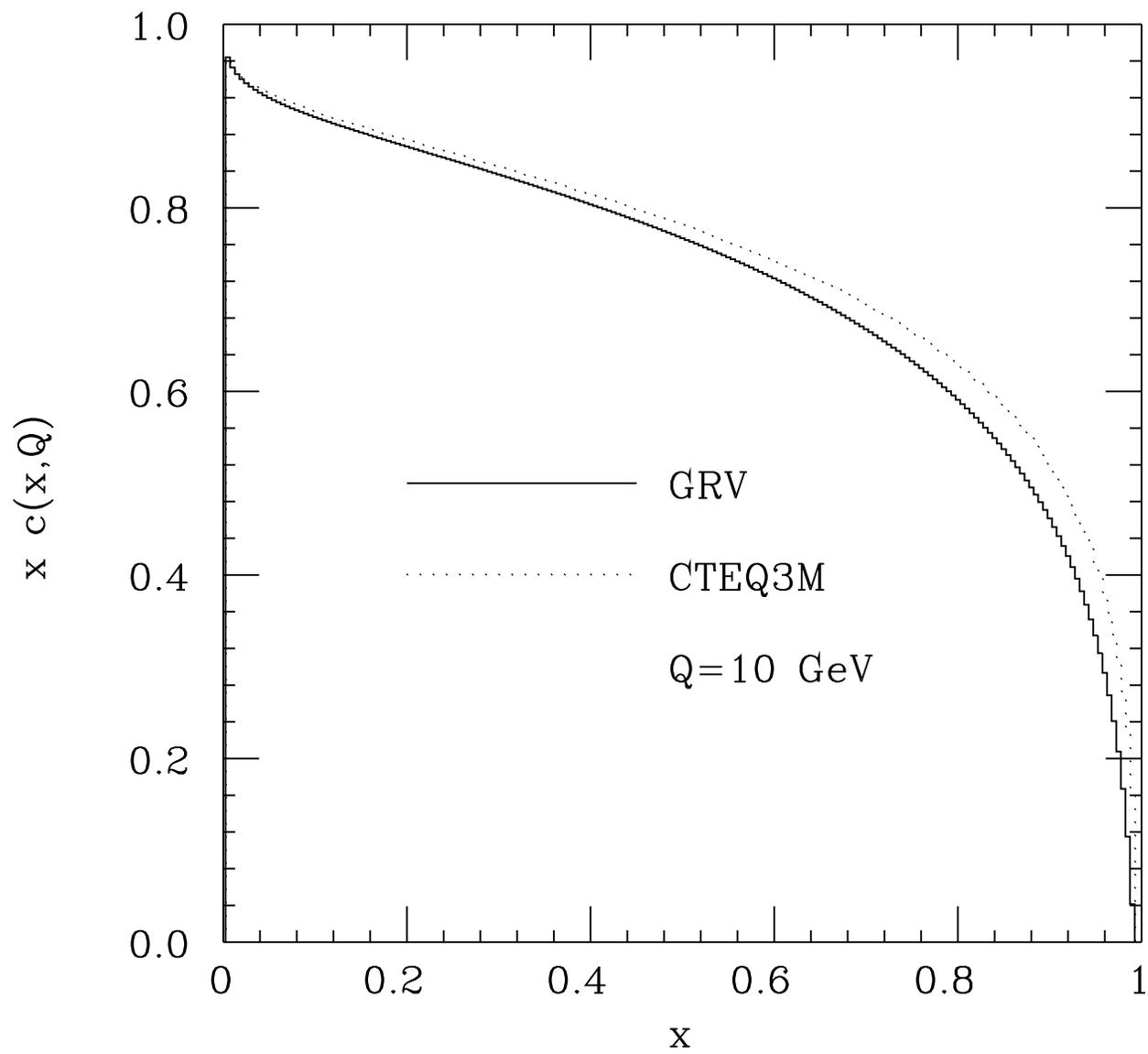
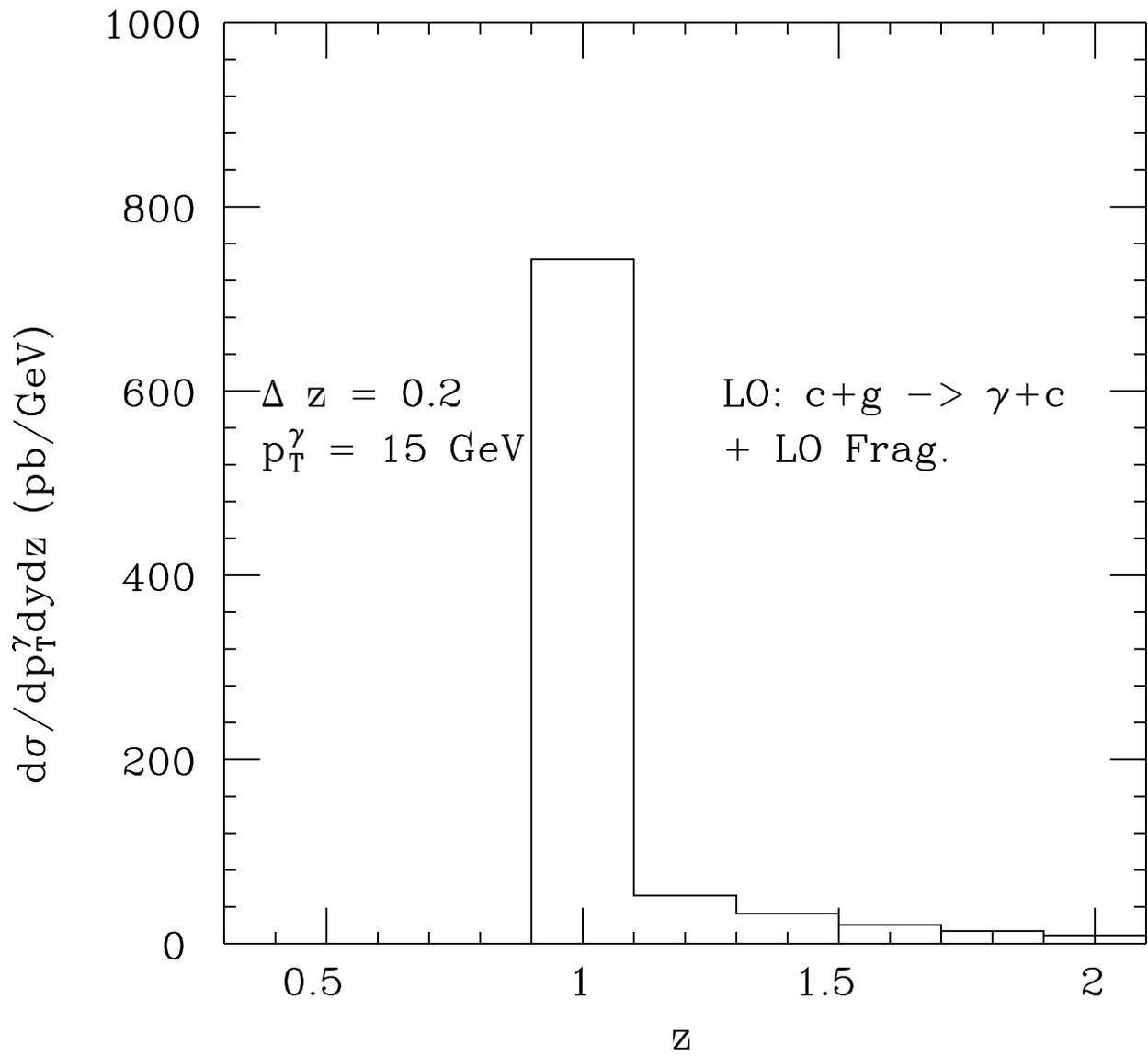
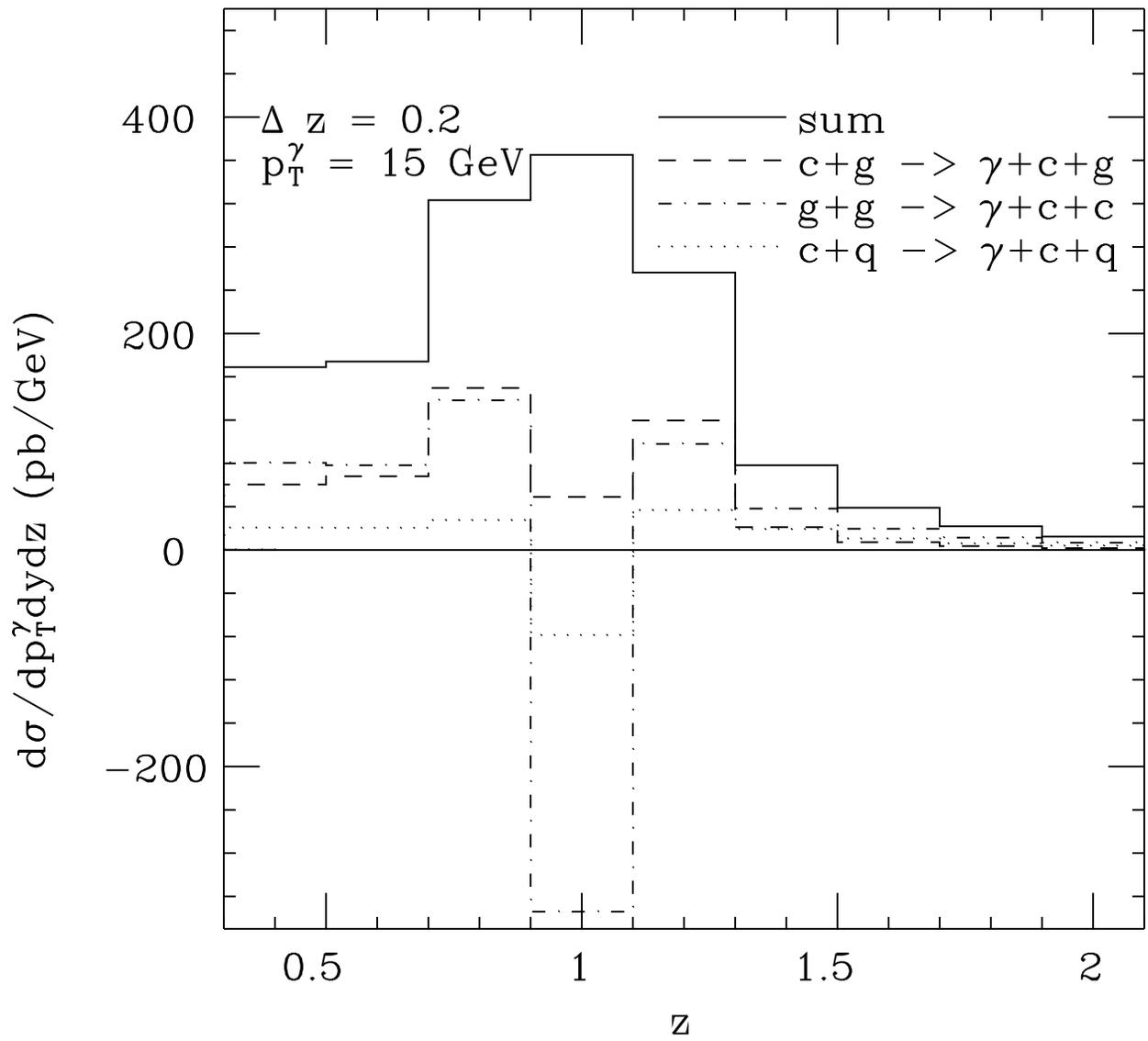
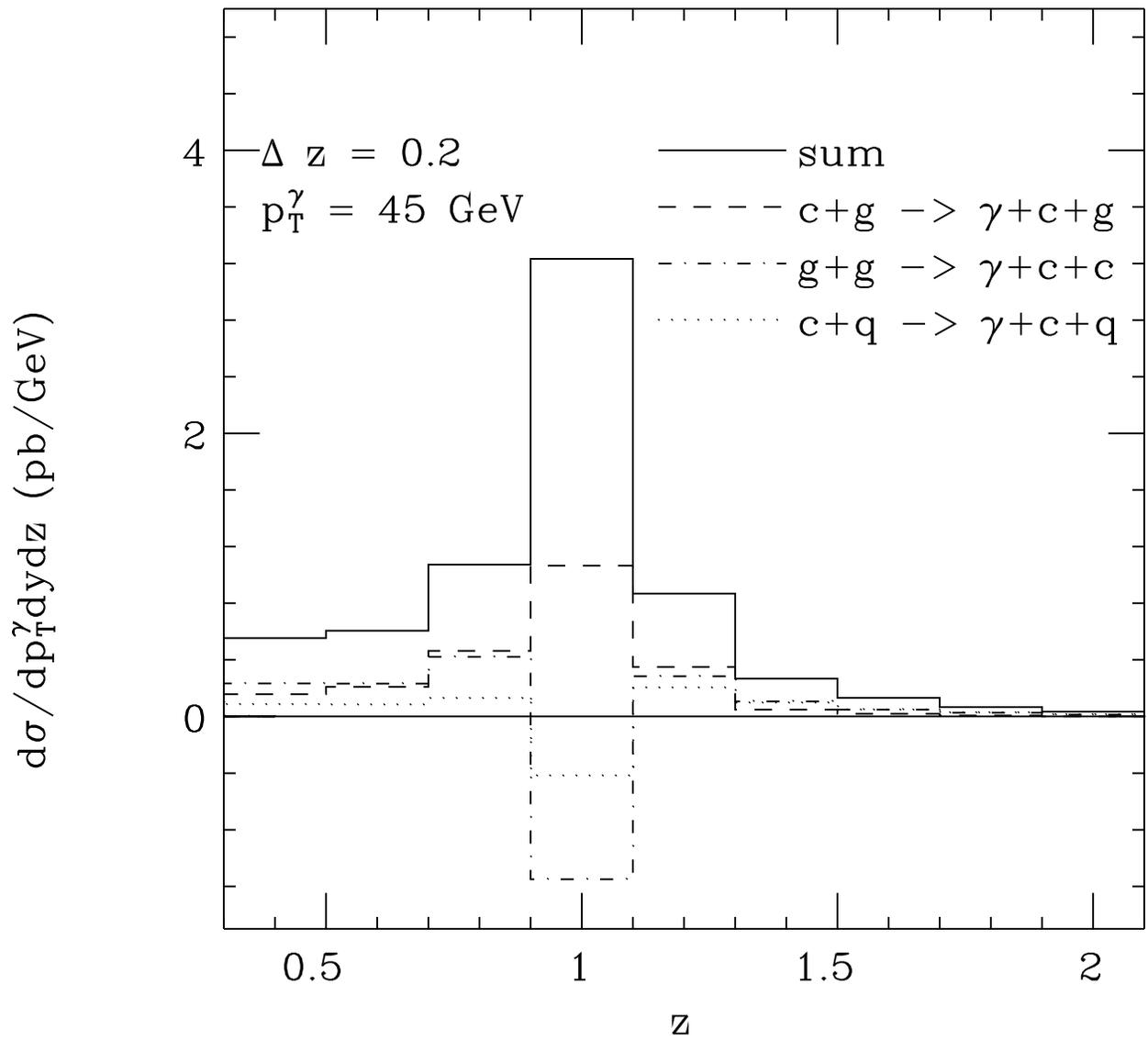
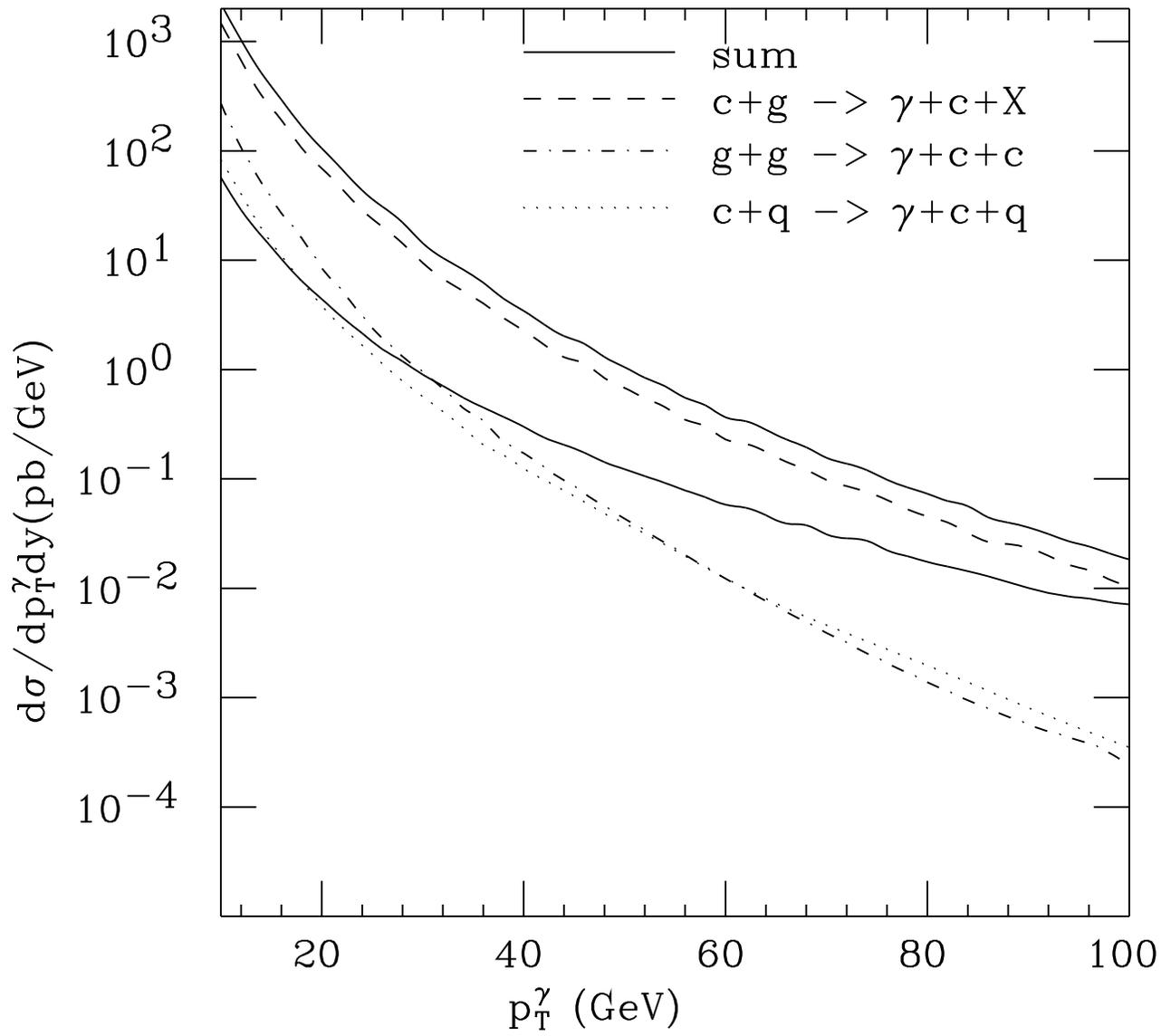
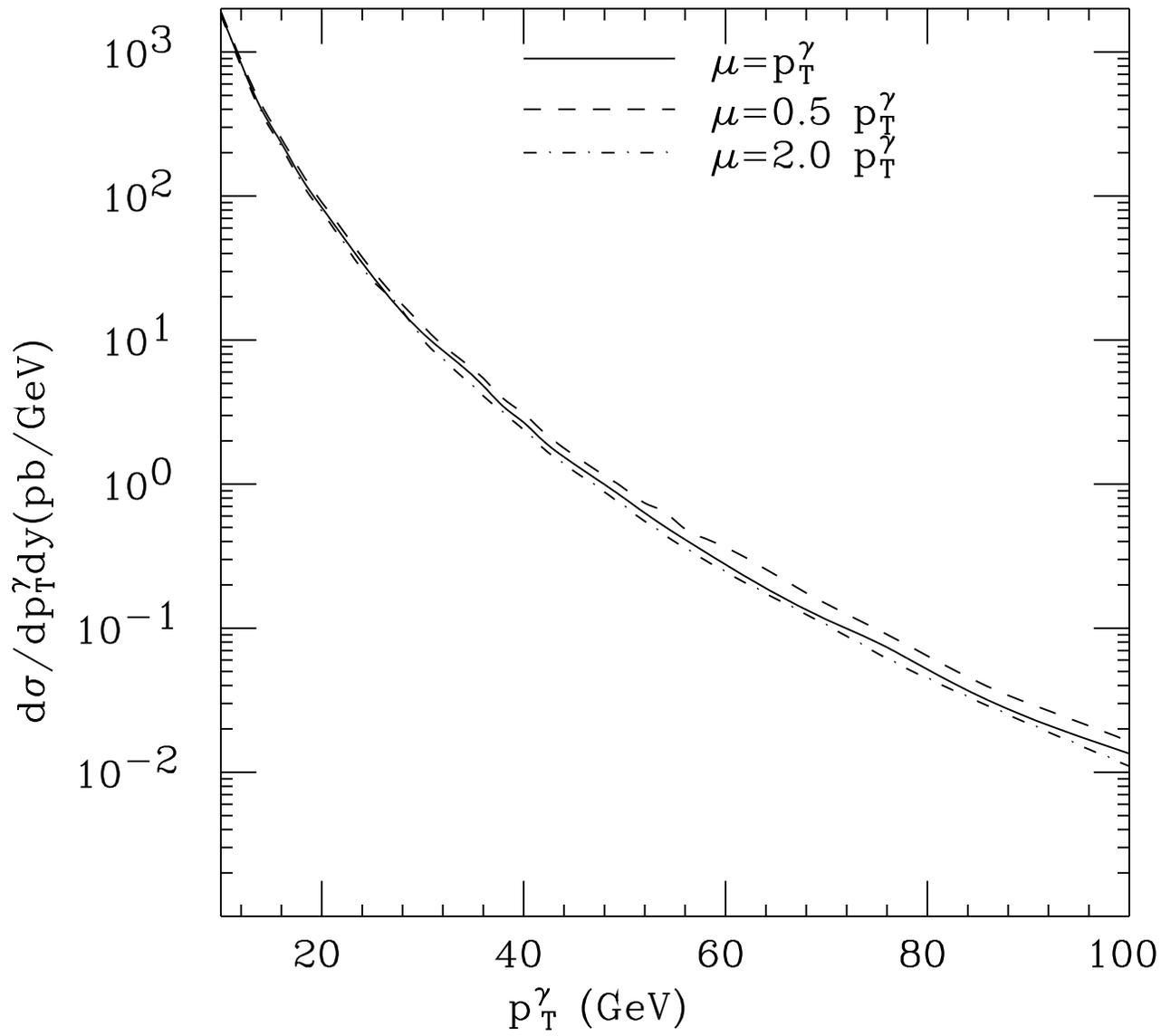
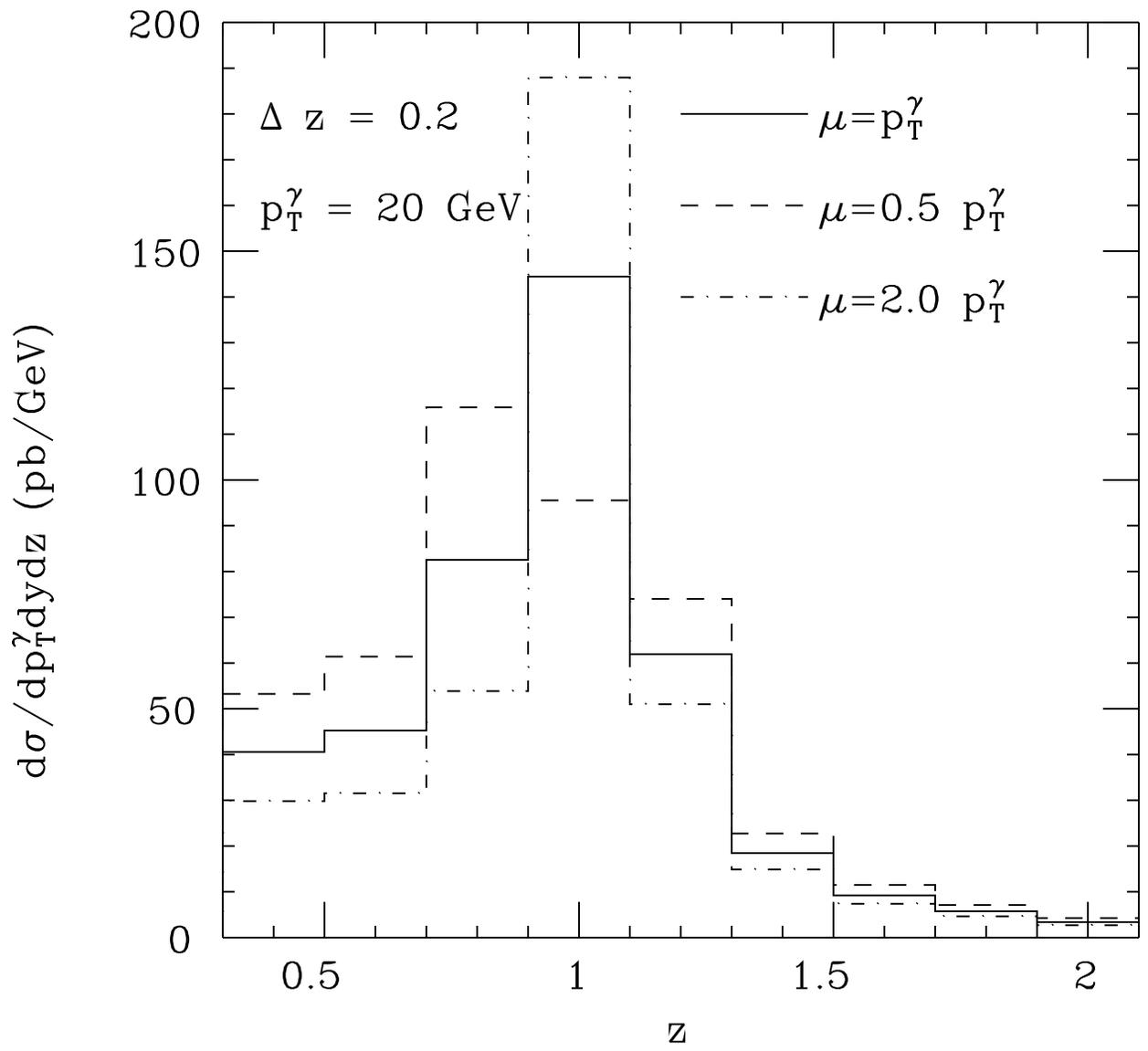
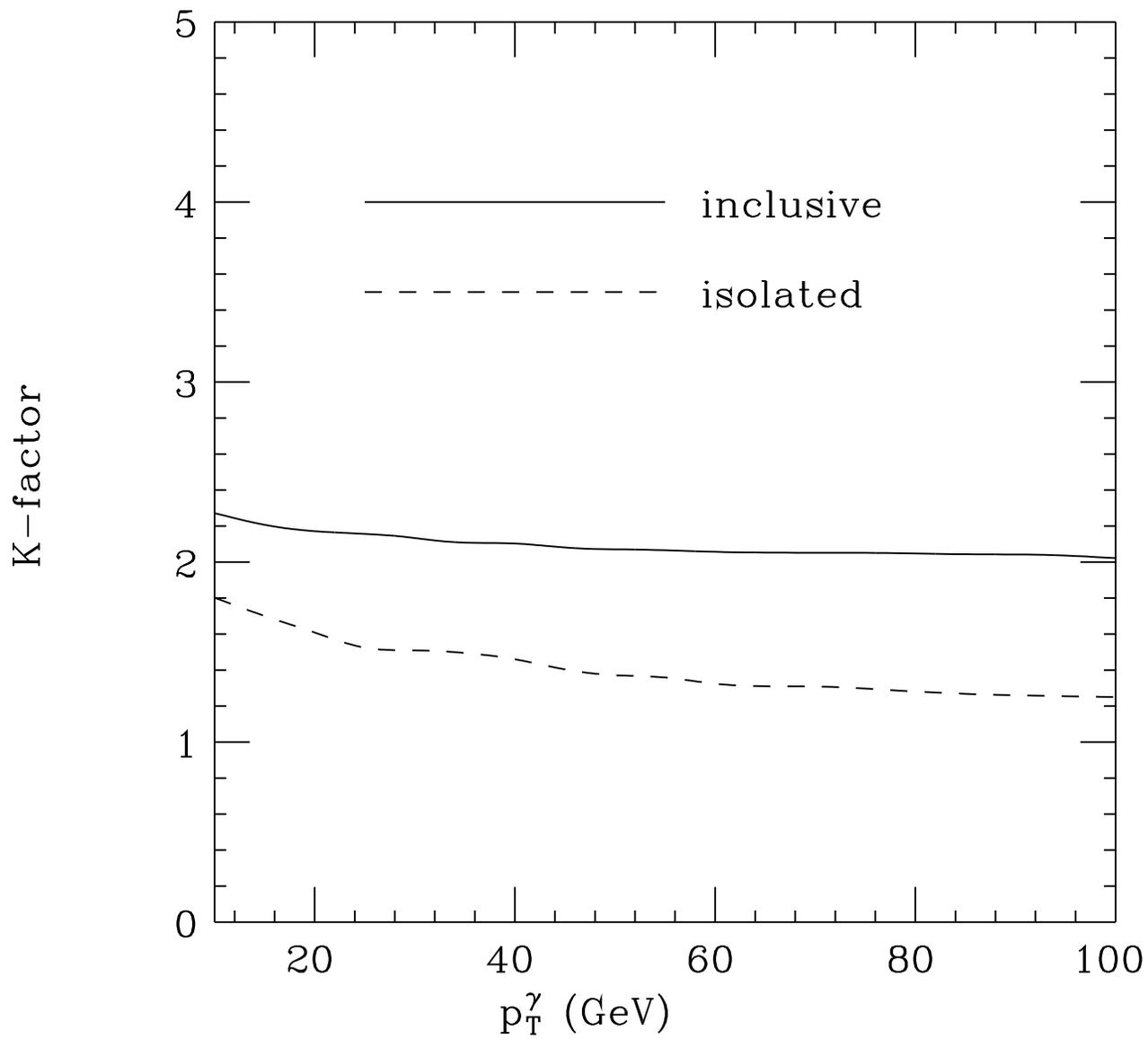
\end{document}